\documentclass[a4paper,11pt]{article}
\pdfoutput=1
\usepackage{jcappub}
\usepackage{inputenc}
\usepackage{amsmath}
\usepackage{amsfonts}
\usepackage{amssymb}
\usepackage{graphicx}
\usepackage{dcolumn}
\usepackage{caption}
\usepackage{float}
\usepackage{bm}
\usepackage{latexsym}
\usepackage[sort&compress]{natbib}
\usepackage[normalem]{ulem}
\usepackage{graphicx}
\usepackage{subcaption}
\usepackage[colorlinks=true]{hyperref}
\usepackage{multirow}
\usepackage{array}
\usepackage{mathtools}

\newcommand{\Ne}{n_{\rm e}}
\newcommand{\NH}{n_{\rm H}}

\newcommand{\op}{\omega_{\rm p}}

\newcommand{\me}{m_{\rm e}}

\newcommand{\mg}{m_{\rm \gamma}}
\newcommand{\ma}{m_{\rm a}}
\newcommand{\Dosc}{\Delta_{\rm osc}}

\newcommand{\De}{\Delta_{\rm e}}

\newcommand{\Da}{\Delta_{\rm a}}
\newcommand{\Dagax}{\Delta^x_{\rm \gamma a}}
\newcommand{\Dagay}{\Delta^y_{\rm \gamma a}}

\newcommand{\Df}{\Delta_{\rm f}}

\newcommand{\Pga}{P({\rm \gamma \rightarrow a})}
\newcommand{\gammaa}{{\rm \gamma a}}
\newcommand{\gammaad}{{\rm \gamma_{ad}}}

\graphicspath{ {figures/} }

\title{A new probe of Axion-Like Particles: CMB polarization distortions due to cluster magnetic fields}
\author[a,b,c] {Suvodip Mukherjee,}\emailAdd{mukherje@iap.fr}
\author[c, d]{David N. Spergel,}\emailAdd{dspergel@flatironinstitute.org}
\author[e] {Rishi Khatri}\emailAdd{khatri@theory.tifr.res.in}
\author[a,b,c,d]{and Benjamin D. Wandelt}\emailAdd{bwandelt@iap.fr}
\affiliation[a]{Institut d'Astrophysique de Paris\\ 98bis Boulevard Arago, 75014 Paris, France}
\affiliation[b]{Sorbonne Universites, Institut Lagrange de Paris \\ 98 bis Boulevard Arago, 75014 Paris, France}
\affiliation[c]{Center for Computational Astrophysics, Flatiron Institute, \\ 162 5th Avenue, 10010, New York, NY, USA}
\affiliation[d]{Dept. of Astrophysical Sciences, Princeton University, Princeton, NJ 08544, USA}
\affiliation[e]{Tata Institute of Fundamental Research\\ Homi Bhabha Road, Mumbai, 400005, India}
\date{\today}

\abstract{We propose using the upcoming Cosmic Microwave Background (CMB) ground based experiments to detect the signal of ALPs (Axion like particles) interacting with magnetic fields in galaxy clusters. The conversion between CMB photons and ALPs in the presence of the cluster magnetic field can cause a polarized spectral distortion in the CMB around a galaxy cluster. The strength of the signal depends upon the redshift of the galaxy cluster and will exhibit a distinctive spatial profile around it depending upon the structure of electron density and magnetic field. This distortion produces a different shape from the other known spectral distortions like $y$-type and $\mu$-type and hence are separable from the multi-frequency CMB observation. The spectrum is close to kinematic Sunyaev-Zeldovich (kSZ) signal but can be separated from it using the polarization information. For the future ground-based CMB experiments such as Simons Observatory and CMB-S4, we estimate the measurability of this signal in the presence of foreground contamination, instrument noise and CMB anisotropies. This new avenue can probe the photon-ALP coupling over the ALP mass range from $10^{-13}$ eV to $10^{-12}$ eV with two orders of magnitude better accuracy from CMB-S4 than the current existing bounds.} 
\begin{document}
\maketitle

\pagenumbering{arabic}
\thispagestyle{plain}
\markright{}

\section{Introduction}
 Since its discovery in 1965, the Cosmic Microwave Background  (CMB) has been a powerful probe of the Universe over a large redshift range $0\leq z \lesssim 2 \times 10^6$ \cite{sz1970, 1982A&A...107...39D, PhysRevD.48.485, Khatri:2012tw}. Different cosmic epochs leave their imprints in the spatial fluctuations of the temperature and polarization field of the CMB and in the CMB  spectral distortions. The distortions in CMB blackbody are expected from several effects that include Sunyaev-Zeldovich ($y$-distortions or thermal SZ (tSZ)) \cite{zeldovich,sz1970}, $\mu$-type \cite{1994ApJ...430L...5H, Chluba:2012gq},  $i$- type (or residual $r$-type) \cite{Khatri:2012tv, Khatri:2012rt,Khatri:2012tw}, relativistic SZ \cite{Hill:2015tqa}, recombination lines \cite{1968ZhETF..55..278Z, 2009AN....330..657S}, and  $\alpha$-type (axions) \cite{Mukherjee:2018oeb,Mukherjee:2018zzg} 
  The spectral distortions signal can also be polarized due to resonant conversion between CMB photons and ALPs \cite{Mukherjee:2018oeb} in the presence of a medium and external magnetic fields.  The CMB-ALPs conversion depends upon the coupling strength $g_{\gamma \gamma a}$ (which we write as $g_{\gamma a}$) between photons and ALPs as well as how slowly (adiabatic) the medium changes along the photon geodesic. The spectral distortions due to ALPs can also arise from  non-resonant conversion and produce spatially fluctuating unpolarized spectral distortions  \cite{sikivie, raffelt_1, anselm,raffelt, Vysotsky:1978dc, Berezhiani:1992rk, Jain:2002vx, Vogel:2017fmc, PhysRevD.76.121301, PhysRevD.84.105030, PhysRevD.86.085036, Schlederer:2015jwa,Mukherjee:2018oeb,Mukherjee:2018zzg}. 
   
Several CMB experiments have explored the spatially varying part of the blackbody and non-blackbody (particularly y-distortions) over the last three decades from space, balloon and ground platforms \footnote{A comprehensive list of the CMB experiments can be found in this website  \url{https://lambda.gsfc.nasa.gov/product/expt/}}. Along with the unprecedented measurement of the temperature and polarization anisotropy by WMAP \cite{2013ApJS..208...20B} and Planck \cite{2016A&A...594A...1P}, ground-based experiments such as ACT (Atacama Cosmology Telescope) \cite{Hasselfield:2013wf} and SPT (South-Pole Telescope) \cite{Bleem:2014iim} have also explored the y-distortion signals from galaxy clusters \cite{Hill:2015tqa, Emami:2015xqa, Ade:2015gva, Staniszewski:2008ma, 2010A&A...518L..16Z, 2013A&A...550A.134P, Khatri:2015jxa}.
  However, the spatially non-varying spectrum of the CMB has not been explored after   COBE-FIRAS (Cosmic Background Explorer-Far Infrared Absolute Spectrophotometer) experiment \cite{firas,firasan,fm2002,2009ApJ...707..916F} which had imposed an upper bound on $y$-distortions and $\mu$-distortions  of $15 \times 10^{-6}$ and $9 \times 10^{-5}$ respectively at $95\%$ C.L.  
  Current and near-term ground-based CMB experiments (such as  Adv-ACT \cite{Henderson:2015nzj}, SPT-3G \cite{Benson:2014qhw}, Simons Array \cite{Stebor:2016hgt}, Simons Observatory (SO) \cite{Ade:2018sbj}, and  CMB-S4 \cite{Abazajian:2016yjj, 2019arXiv190704473A}) and the upcoming space-based CMB experiment such as LiteBIRD \cite{2014JLTP..176..733M}, and several proposed CMB experiments (such as CMB-Bharat \footnote{Proposal for a space-based CMB mission is  submitted to Indian Space Research Organization (ISRO).}, PRISTINE (Polarized Radiation Interferometer for Spectral
disTortions and INflation Exploration) and PICO (Probe of Inflation and Cosmic Origins) \cite{2018arXiv180801369Y, 2019arXiv190210541H}) plan to  explore both blackbody and non-blackbody part of CMB with much higher instrument sensitivity and finer angular resolutions than the ongoing surveys.

In this paper, we explore the signatures of the polarized spectral distortion of CMB originating from galaxy clusters due to resonant photon-ALP conversion. Galaxy clusters possess magnetic fields of the order of $\mu$-Gauss ($\mu$G) as evident from multiple observations \cite{doi:10.1146/annurev.astro.40.060401.093852, Boehringer:2016kqe}. As a result, CMB photons while passing through galaxy clusters can undergo conversion into ALPs. The component of the CMB photons with polarization vector parallel to the local magnetic field direction gets converted into ALPs while the other polarization state remains unaltered. As a result, every conversion of CMB photons into ALPs leads to a polarized distortion in the blackbody spectrum of the CMB. For non-resonant conversion, these distortions take place everywhere along the line of sight and if the magnetic field is not coherent but turbulent, both polarizations of the CMB photons will be converted to ALPs with equal probability leading to an unpolarized distortion signal in the clusters. The typical strength of this distortion is expected to be small due to high electron densities present in these systems \cite{Mukherjee:2018oeb}. A detailed study of this effect is performed for Milky Way galaxy in  \cite{Mukherjee:2018oeb} and these results are also  applicable to galaxy clusters. However, the polarization signal can be preserved for the resonant conversion case which takes place only at a specific spatial location where the ALP mass is equal to the effective mass of photons. The polarized photon-ALP signal has a radial profile that depends upon the spatial structure of electron density and magnetic field in the cluster.

The paper is organized as follows.  We review the basic formalism of resonant photon-ALP conversion in Sec.~\ref{ALP-formalism}. In Sec.~\ref{ALP-cluster} and Sec.~\ref{ALP-contamination} respectively, we describe the signal from a single cluster and simulate a realistic sky signal by adding galactic and extragalactic contamination. The forecast for measurability of this signal from future CMB experiments such as Simons Observatory (SO) \cite{Ade:2018sbj} and CMB-S4 \cite{Abazajian:2016yjj, 2019arXiv190704473A}  are presented in Sec.~\ref{ALP-forecast}. Finally,  future prospects and our conclusions are presented in Sec.~\ref{ALP-conclusion}.

\section{Resonant photon-ALP conversion signal: Formalism}\label{ALP-formalism}
Photon to ALP conversion in presence of magnetic field and electron density can be expressed in the relativistic limit by the following coupled system of equations \cite{raffeltbook} \footnote{Analogous to the Mikheyev-Smirnov-Wolfenstein (MSW) effect \cite{Wolfenstein:1977ue,Mikheev:1986gs,Mikheev:1986if} in neutrino oscillations}

\begin{align}
\left(\omega+\left(
\begin{array}{ccc}
\De & \Df & \Dagax\\
\Df & \De & \Dagay\\
\Dagax & \Dagay & \Da\\
\end{array}
\right)+i\partial_z\right)
\left(
\begin{array}{c}
A_{x}\\
A_{y}\\
a\\
\end{array}
\right)=0,\label{eq-evolution}
\end{align}
where ($A_x, \, A_y$) are two polarizations of photons propagating in the z-direction, $a$ is the axion field and the entries in the coupling matrix are defined as 
\begin{align}
\begin{split}
\left(\frac{\Delta^{ i}_{\gammaa}}{{\rm{Mpc^{-1}}}}\right) &\equiv \frac{g_{\gammaa} |B_{i}|}{2} = 15.2
\left(\frac{g_{\gammaa}}{10^{-11}{\rm Gev}^{-1}}\right)\left(\frac{B_{i}}{\mu{\rm G}}\right),\\
\left(\frac{{\Da }}{{\rm{Mpc^{-1}}}}\right) &\equiv - \frac{m^2_a}{2\nu}= -1.9\times
10^4\left(\frac{\ma}{10^{-14}{\rm eV}}\right)\left(\frac{100~ {\rm
      GHz}}{\nu}\right),\\
        \left(\frac{\De}{{\rm{Mpc^{-1}}}}\right) &\approx   \frac{\op^2}{2\nu}
\left[-1 +7.3\times 10^{-3}\frac{\NH}{\Ne}  \left(\frac{\omega}{{\rm
        eV}}\right)^2\right] \\
&=-2.6\times
10^6 \left(\frac{\Ne}{10^{-5}\,{\rm cm^{-3}}}\right) \left(\frac{100~{\rm
    GHz}}{\nu}\right) \times \left[1 -7.3\times 10^{-3}\frac{\NH}{\Ne}  \left(\frac{\omega}{{\rm
        eV}}\right)^2\right],\,\\
\bigg(\frac{\Delta_f}{\text{Mpc}^{-1}}\bigg)&= (n_+-n_-)\nu\, = 7.3 \times 10^{-3}  \bigg(\frac{n_e}{10^{-3}\, \text{cm}^{-3}}\bigg) \bigg(\frac{B_{||}}{\mu G}\bigg) \bigg(\frac{100\,  \text{GHz}}{\nu}\bigg)^2,
\end{split}
\end{align}
where $\op^2=4\pi \alpha\Ne/(\me)$ is the plasma frequency, $m_e$ is the electron mass, $n_e$ is the electron density, $\alpha$ is the fine structure constant, $m_a$ is the mass of ALP, $B_{i}$ is the magnetic field along the direction $\hat i$ (where $i$ can be $x$ or $y$), $n_H$ is the density of the hydrogen atoms and $\Df$ gives rise to the Faraday rotation written in terms of the difference in the refractive index for left ($n_-$) and right ($n_+$) component circularly polarized light \cite{1979rpa..book.....R, Takada:2001bm}. Typical lines of sight through a galaxy cluster  will undergo a Faraday rotation of the polarization plane by about an arc-min for the frequency of CMB photons $\nu\geq90$ GHz. This is a minor rotation of the plane of polarization in comparison to the resolution of the instrument beam \footnote{See Table \ref{tab:so} and Table \ref{tab:s4}} and hence we ignore the effect from Faraday rotation in our analysis. We can reduce Eq.~\eqref{eq-evolution} to a two flavour system by choosing the $x$-axis along the magnetic field, so that $\Delta^y_{\gamma a}= 0$ and $A_y$ is decoupled from the three flavour system consisting of $A_x$ and $a$ from the other two eigenstates. The dispersion relation that defines the two eigenstates of the photon-ALP system is then given by \cite{Raffelt:1996wa} 
\begin{align}
2\omega(\omega-k)&=-\omega\left(\De+\Da\right)\pm \omega\Dosc\nonumber\\
&=\frac{\ma^2+\mg^2}{2}\pm\left[\left(\frac{\ma^2-\mg^2}{2}\right)^2+\omega^2g_{\gammaa}^2B_{\rm T}^2\right]^{1/2},
\end{align}
where $\Dosc^2 = (\Da  - \De )^2 +4\Delta^2_{\gammaa}$ and $m_\gamma$ is the effective mass of the photon defined as $m^2_\gamma= \frac{4\pi\alpha n_e}{m_e}$. The detailed calculations of resonant and non-resonant photon-ALP conversion can be found in \cite{Raffelt:1996wa, Mukherjee:2018oeb,Schlederer:2015jwa}.  In the remainder of this section, we only overview the formalism of resonant-conversion between photons and ALPs which is relevant for this paper.  {The contribution from the non-resonant photon-ALPs conversion will be explored in a future work.} 

Resonant photon-ALP conversion happens when the ALP mass becomes equal to the photon effective mass in the plasma, i.e. $m_a= m_\gamma$. As a result the mixing angle ($\theta=\frac{1}{2}\cos^{-1}(\frac{\Da-\De}{\Dosc})$) becomes $\pi/4$. In the galaxy cluster, the electron density and magnetic field change with spatial position in the cluster. So, the important quantity to compare is the length scale over which the electron density and magnetic field vary with the oscillation length scale of photon-ALP conversion $\Dosc^{-1}$. The ratio of these two quantities defines the adiabaticity  parameter 
\begin{align}
\gammaad 
&=\left|\frac{\Dosc^2}{\nabla\De}\right|_{\text{at resonance}}, \\&= \left|\frac{2g^2_{\gammaa}B^2\nu}{\nabla\omega^2_p}\right|_{\text{at resonance}} ,\label{Eq:gad}
\end{align}
where $\nabla$ denotes spatial gradient. For $\gamma_{ad} >>1$, we are in the adiabatic limit resulting in complete conversion of photons to axions or vice versa when passing a single resonance. The value of $\gammaad<<1$ denotes the non-adiabatic limit, which implies that the length scale over which oscillation is happening is much larger than the physical scale over which the electron density is varying.  For a locally (where $m_a=m_\gamma$) linearly varying electron density, the transition probability from initial eigenstate $|\psi_i(0)\rangle$ to
  final eigenstate $|\psi_j(1)\rangle$ after crossing one resonance can be written as \cite{kp1989,Fogli:2003dw}
\begin{align}
\left|\langle\psi_i(1)|\psi_j(0)\rangle\right|^2=\left(
\begin{array}{cc}
1-p & p\\
p & 1-p\\
\end{array}
\right),
\end{align}
where $p$ is given by the Landau-Zener formula \cite{landau1932,zener1932,stueckelberg1932,kp1989,Raffelt:1996wa} 
\begin{align}\label{trans-prob}
p&=e^{-\pi \gammaad/2},\\
&\approx 1-\frac{\pi \gammaad}{2}\, \text{for $\gammaad <<1$},
\end{align}
in the non-adiabatic limit. The initial and final eigenstates are superposition of both photons and ALPs determined by the mixing angle $\theta$ and the corresponding conversion probability from photon to ALP can be calculated as  \cite{Raffelt:1996wa, Mukherjee:2018oeb, {Schlederer:2015jwa}}
\begin{align}
P(\gamma \rightarrow a) =\frac{1}{2}\left(1-(1-2p)\cos 2\theta_0\cos 2\theta_d\right),\label{Eq:level}
\end{align}
where $\theta_0$ and $\theta_d$ are the initial mixing angle when the photon is emitted and the mixing angle at the detection, respectively. Assuming that emission and detection happens far from resonance gives $\theta_0 \approx \theta_d \approx 0$ and $P(\gamma \rightarrow a) = \frac{1}{2} (1- (1-2p))$. 
In the case of multiple ($N$) resonances, the transition probability from initial eigenstate to final eigenstate can be written as \cite{kp1989,Fogli:2003dw}
\begin{align}
\left|\langle\psi_i(N)|\psi_j(0)\rangle\right|^2&=
\left|\langle\psi_i(N)|
  \cdots|\psi_k(a)\rangle\langle\psi_k(a)|\psi_l(a-1)\rangle\langle\psi_l(a-1)| \cdots \psi_j(0)\rangle\right|^2,\nonumber\\
 &= \prod_{a=1}^{N}\left(
\begin{array}{cc}
1-p_a & p_a\\
p_a & 1-p_a\\
\end{array}
\right)
\equiv
\left(
\begin{array}{cc}
1-p & p\\
p & 1-p\\
\end{array}
\right),
\end{align}
 where $p$ is the net level crossing probability which can be written in terms of the level crossing probability $p_a$ at each $a^{th}$ resonance as
\begin{align}
p=\frac{1}{2}\left(1-\prod_{a=1}^{N}\left(1-2p_a\right)\right).\label{Eq:levelp}
\end{align}
In the above calculation we have ignored the interference between different resonances and treated the level crossing probabilities as classical probabilities.  Eq.~\eqref{Eq:levelp} along with  Eq.~\eqref{Eq:level} determines the conversion probability from photons to ALPs after $N$ resonances. Given a photon produced in the high electron density region and detected at a high (or low) electron density region, uniquely determines whether the resonant conversion is even (or odd). There remains no additional  freedom to choose both $N$ as even (or odd) and the electron density at the detection independently. Fixing one of them naturally fixes the other one. As a result we can write Eq. \ref{Eq:level} as \cite{Mukherjee:2018oeb}
\begin{align}\label{net-prob}
\Pga= \begin{array}{cc}
p &: N~{\rm even},\\
1-p &:N~{\rm odd}.
\end{array}
\end{align}

With the probability of transition given by Eq.~\eqref{trans-prob}, we can write Eq.~\eqref{net-prob} after retaining only the leading order terms in $\gamma_{ad}$ as 
 \begin{align}\label{net-prob-2}
\Pga= \begin{array}{cc}
\frac{\pi \gammaad}{2} &: N=1,\\
\frac{\pi (\gammaad_1 + \gammaad_2)}{2} &:N=2.\\
\end{array}
\end{align}

The preceding description of resonant photon-ALP conversion is general; it can take place in different scenarios whenever the electron density changes along the photon path and there is an external magnetic field present transverse to the photon propagation direction. Examples of physical effects due to resonant conversions include the cosmic evolution of the electron density in the Universe \cite{Jaeckel:2008fi, Mirizzi:2009nq, Mukherjee:2018oeb}, giving a monopole spectral distortion in the CMB blackbody; or  resonant conversion in the Milky Way \cite{Mukherjee:2018oeb}.  In this  paper we focus on the resonant photon-ALP conversion in galaxy clusters.

\section{Resonant photon-ALPs conversion in a galaxy cluster}\label{ALP-cluster}
\begin{figure}
      \centering
     \includegraphics[trim={0 0 0 0cm}, clip, width=0.8\textwidth]{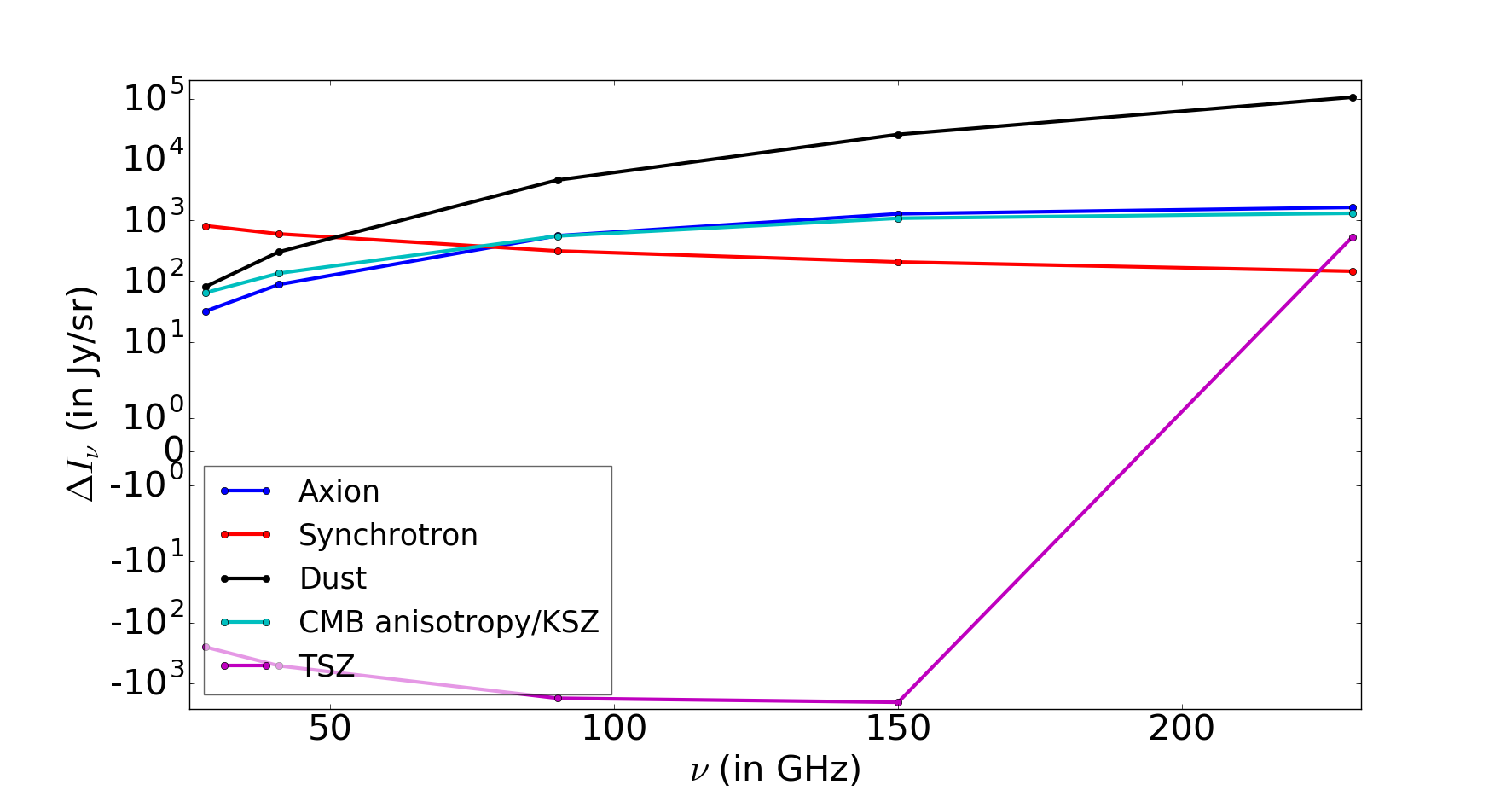}
          \captionsetup{singlelinecheck=on,justification=raggedright}
     \caption{Comparison of the photon-ALP signal to standard model spectral distortions. While the kSZ and the photon-ALP distortions have nearly degenerate distortions, they can be distinguished because the photon-ALP signal is polarized and the kSZ is not. Shown are photon-ALP distortions (for $g_\gammaa = 10^{-12}$ GeV$^{-1}$, and $m_a= 10^{-13}$ eV),  synchrotron (for the amplitude $A_{sync}= 288$ Jy/sr and spectral index, $\alpha_s= -0.82$), dust (for the amplitude $A_{dust}= 1.36 \times 10^6$ Jy/sr and dust temperature $T_d= 21$K, dust spectral index $\beta_d=1.53$), CMB anisotropy/kSZ (for the amplitude $\Delta T= 10^{-6}$K) and tSZ (for the amplitude $y= 2\times 10^{-6}$) are shown for the frequency channels of the proposed CMB-S4 project. \cite{Abazajian:2016yjj, 2019arXiv190704473A}.  }\label{Fig:Inu-nu}
     \end{figure}

The polarized spectral distortion in presence of the magnetic field from galactic clusters leaves its imprint in the $Q$ and $U$ Stokes parameters of the CMB polarization map ( {see \cite{Hu:1997hv} for a review}). The cluster locations are readily identifiable using the tSZ (Thermal Sunyaev-Zeldovich) signal.  We see in Fig.~\ref{Fig:Inu-nu} that the spectrum of resonant photon-ALP distortion is very close to the kSZ (kinematic SZ) signal. Therefore polarization information is crucial in distinguishing between the two signals. The kSZ effect does not have the same profile, especially for small axion masses. That can help to distinguish them even in the case of partial (or complete depolarization). \footnote{A previous study \cite{Schlederer:2015jwa} addressed the photon-ALP signal from the Coma and Hydra clusters without exploiting the unique polarization signature to disentangle the signal. While this study considered the tSZ effect as foreground, it ignored the kSZ  effect which would be the most worrisome contamination in their analysis.} The scattering by the electrons in the galaxy clusters of the local (at the redshift of the cluster) CMB quadrupole can generate a small polarized signal and  an estimate of this signal is discussed in Sec.~\ref{amp_alps}. 
     
The distortion in the CMB due to the photon-ALP conversion (which we refer as $\alpha$-type distortion) is given by 
\begin{align}
\begin{split}
\Delta I_{\nu} =& \Pga I^{BB}_{\nu} ,\\
&= \alpha_{\gammaa}\,fI^{BB}_{\nu}\label{Eq:bb-dist}
\end{split}
\end{align}
where $ I^{BB}_{\nu}=\left(\frac{2h\nu^3}{c^2}\right)\frac{1}{(e^f-1)}$ is the intensity of the blackbody spectrum of CMB, with $f= h\nu/k_BT_{CMB}$ and $T_{CMB}= 2.7255$K and $h$, $c$ and $k_B$ are Planck's constant, speed of light and Boltzmann constant respectively. $\alpha_{\gammaa}\equiv \Pga/f$ is the dimensionless ALPs distortion parameter.

In order to calculate the CMB distortion from a single galaxy cluster, we need a model of electron density and magnetic fields inside the cluster. For simplicity, one can consider two extreme cases. Case I: The electron density and magnetic fields inside the cluster are completely stochastic and turbulent. In this case, we will have multiple resonances with arbitrary direction for magnetic field and therefore the polarization direction of CMB photons converting to ALPs. The result will be a completely depolarization of the $\alpha$-type distortion. Since the spectrum is close to the kSZ spectrum, such a distortion will be very hard to separate from kSZ signal. Ref. \cite{Schlederer:2015jwa} studied this case but ignored the denegracy with kSZ effect. Case-II: We will study the second extreme where turbulence is negligible and the magnetic field and electron density are described by a smooth profile as a function of the radius of the cluster. A real cluster would be in between these two extremes \cite{Govoni:2004as,1538-4357-547-2-L111, 2010NatPh...6..520P}. We would expect to reach this smooth limit for clusters in hydrostatic equilibrium.  {In this analysis, we have considered the strength of the magnetic field from the center of the galaxy cluster to follow a radial profile and the direction of the magnetic field is considered to be randomly oriented. The model of electron density and magnetic field considered in this analysis is discussed below. The impact of the variation in the magnetic field radial dependence on the photon-ALPs conversion probability is discussed in Sec. \ref{amp_alps}. A detail study showing the variation of the photon-ALPs signal for different models of electron density and magnetic field will be presented in a future analysis.}

\textbf{Electron density :} Electron density in galaxy clusters has been well studied in  observations \cite{1978A&A....70..677C, Vikhlinin:2005mp, 2017A&A...608A..88B, 2016A&A...590L...1P, McDonald:2013fka,McDonald:2017ypo} as well as from hydrodynamical simulations \cite{Battaglia:2016xbi}.
The large scale spatial structure of electron density in galaxy clusters is well described by the modified $\beta$-electron model \cite{1978A&A....70..677C, Vikhlinin:2005mp, 2017A&A...608A..88B, 2016A&A...590L...1P, McDonald:2013fka,McDonald:2017ypo}.  In this analysis, we consider the following observation motivated modified $\beta$-electron density model \cite{Vikhlinin:2005mp}
\begin{equation}\label{elec-den}
n_pn_e= n_0^2\frac{(r/r_{c1})^{-\alpha}}{((1+ r^2/r^2_{c1})^{3\beta- \alpha/2}) ((1+ r^\gamma/r^\gamma_s)^{\epsilon/\gamma})} + \frac{n^2_{02}}{(1+ r^2/r^2_{c2})^{3\beta_2}},
\end{equation}
where the values of the parameters are chosen as $r_c= 100$ kpc, $n_0= 10^{-3}$ cm$^{-3}$, $r_s= 1000$ kpc, $\alpha= 2$, $\gamma= 3$, $r_{c2}= 10$, $\beta= 0.64$, $\beta_2= 1$, $n_{02}= 10^{-1}$ cm$^{-3}$ from \cite{Vikhlinin:2005mp}.
Recent studies \cite{McDonald:2017ypo}  have shown that the electron density in the cluster are remarkably self-similar over a wide range of redshift $z= 0-1.9$ beyond the inner core of the cluster ($r\, \geq\, 100$kpc), with error-bar about $20\%-40\%$ (See for Fig. 3 in the ref. \cite{McDonald:2017ypo}). 
Variation in the profile of the cluster electron density will directly translate into a variation in the $m_a^2$ and $\nabla n_e$. The variation of  the photon-ALP signal with the variation in  the electron density will be explored in a future work.  

\textbf{Magnetic fields :} There have been several studies attempting to measure the typical strength of the magnetic field in  galaxy clusters using radio observations, synchrotron emission, and Faraday rotation in the clusters \cite{doi:10.1146/annurev.astro.40.060401.093852, Boehringer:2016kqe}. The typical strength obtained from these observations indicates that the magnetic field in the galaxy clusters is of the order of $\mu$G. It also exhibits a dependence on the electron density $B(r)= (n_{e}(r)/n_{0})^\eta$  \cite{2010arXiv1009.1233B}, with the value of $\eta=0.5$ for the Coma cluster \cite{2010arXiv1009.1233B}. We therefore adopt a simple model of the radial profile of the strength of the cluster magnetic field  
\begin{equation}\label{mag-field}
B(r)= \frac{B_0}{\sqrt{(r/r_b)}},
\end{equation}
with $r_b = 10$kpc and $B_0= 3\, \mu$G. In reality,  a turbulent component may be present in addition to  the large scale coherent magnetic field. As long as the magnetic fields are coherent on scales of angular resolution of our telescope, the exact structure of the magnetic fields does not affect our main results. Turbulence on scales smaller than the beam size will however at least partially depolarize the distortion. For this study we will assume that these effects are small and leave a more detailed calculation for future work.

For the resonant photon-ALP conversion, the signal strength along a particular line of sight does not depend upon the complete magnetic field structure, but only depends on the local region where the resonant conversion takes place (i.e. $m_a = m_\gamma$). So, locally within a narrow spatial range, we can consider a uniform value of the magnetic field. 
From the current models of magnetic field \cite{2010arXiv1009.1233B, Boehringer:2016kqe}, we expect the magnetic field of the order $0.1\,\mu  G$ on the Mpc scales. In our analysis, we have taken the direction of the magnetic field in the cluster as random  {and its strength varies} with radius according to Eq.~\eqref{mag-field}.  

\subsection{Polarized spectral distortion of CMB around a single galaxy cluster}\label{ALP:singlecluster}
\subsubsection{Spatial Shape  of the ALPs-distortion}
CMB photons passing through the galaxy clusters can undergo resonant conversion into ALPs in the spatial regions where the condition $m_a = m_\gamma$ is satisfied. In the case of multiple resonances, the total conversion is given by  Eq.~\eqref{net-prob-2}. For a spherically symmetric monotonically changing electron density profile, photons can convert into ALPs of a particular mass $m_a$ at most twice: (i) entering and (ii) leaving the galaxy cluster. As a result, in the frame of the galaxy cluster, the conversion  from photons to ALP of a particular mass will appear as a spherical shell of radius $r_0$ set by the condition $m_a= m_\gamma(r_0)$.

\begin{figure}[H]
\centering
\includegraphics[trim={2.cm 2.cm 2.cm 2.cm}, clip, width=0.6\textwidth]{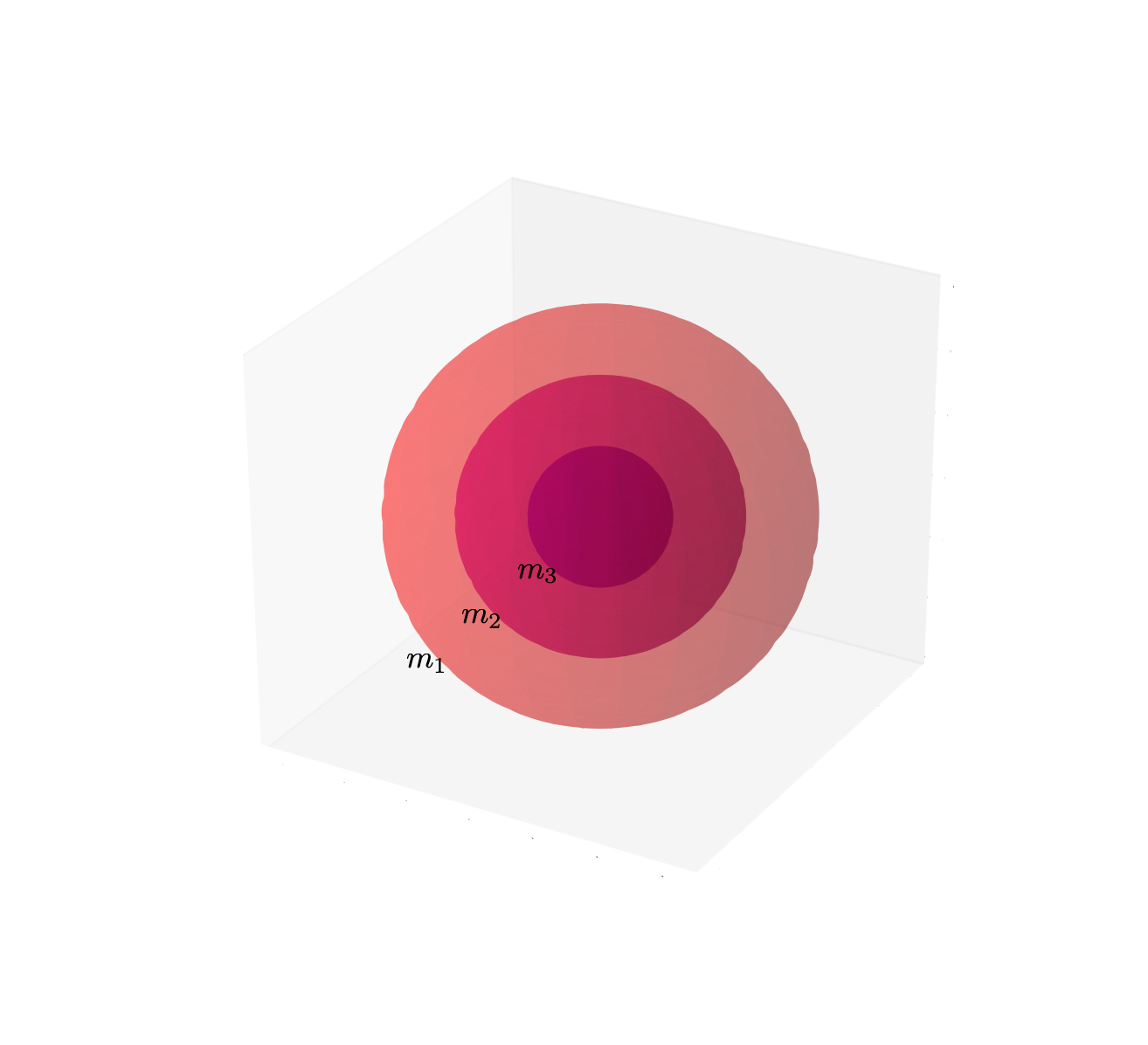}
      \captionsetup{singlelinecheck=on,justification=raggedright}
 \caption{The schematic diagram showing the spatial structure of resonant photon-ALP conversion in the galaxy cluster for a spherically symmetric electron density. For three different ALP masses $m_1$ (light red), $m_2$ (medium red) and $m_3$ (dark red) (with $m_1<m_2<m_3$), resonant conversion happens within a narrow shell at different radii $r_1, r_2, r_3$ respectively with  $r_1>r_2>r_3$. The projection of the $3$-D spherical shell, will appear as a disk in the plane of the sky.}\label{Fig:mass-shell}
\end{figure}

The projection of the $3$-D spherical shell on the two dimensional plane of the sky will appear like a disc as shown in Fig.~\ref{Fig:mass-shell}. 
 For ALPs of different masses, the resonant conversion happens at different radii and hence produces discs of different sizes (See Fig.~\ref{Fig:mass-r-1}).

\subsubsection{Amplitude of the photon-ALP distortion in a galaxy cluster}\label{amp_alps}
The adiabaticity parameter for a galaxy cluster at a redshift $z$ can be written  as
\begin{align}
\begin{split}
\gammaad (r,z)
&= \left|\frac{2g^2_{\gammaa}B^2(r,z)\nu_0(1+z)}{\nabla\omega^2_p (r,z)}\right| ,\\\label{Eq:gad-2}
\end{split}
\end{align}
where $\nu_0$ is the observed frequency of the CMB photons. In this analysis, we have taken  universal magnetic field \cite{Boehringer:2016kqe} and electron density \cite{Vikhlinin:2005mp, 2017A&A...608A..88B, 2016A&A...590L...1P, McDonald:2013fka} profiles of galaxy clusters. Eq.~\eqref{Eq:gad-2} implies that $P(\gamma \rightarrow a)$ around every cluster can exhibit a redshift dependence through the factor $(1+z)$.  The radial profile of the distortion follows the spatial structure of the electron density and magnetic field in the galaxy clusters. The angular size of the distortion will be decided by the ratio of the radius at which resonant photon-ALP conversion is taking place and the angular diameter distance to the galaxy cluster.

\begin{figure}[H]
\begin{subfigure}{1\linewidth}
\centering
\includegraphics[trim={0cm 0 0cm 0cm}, clip, width=0.88\textwidth]{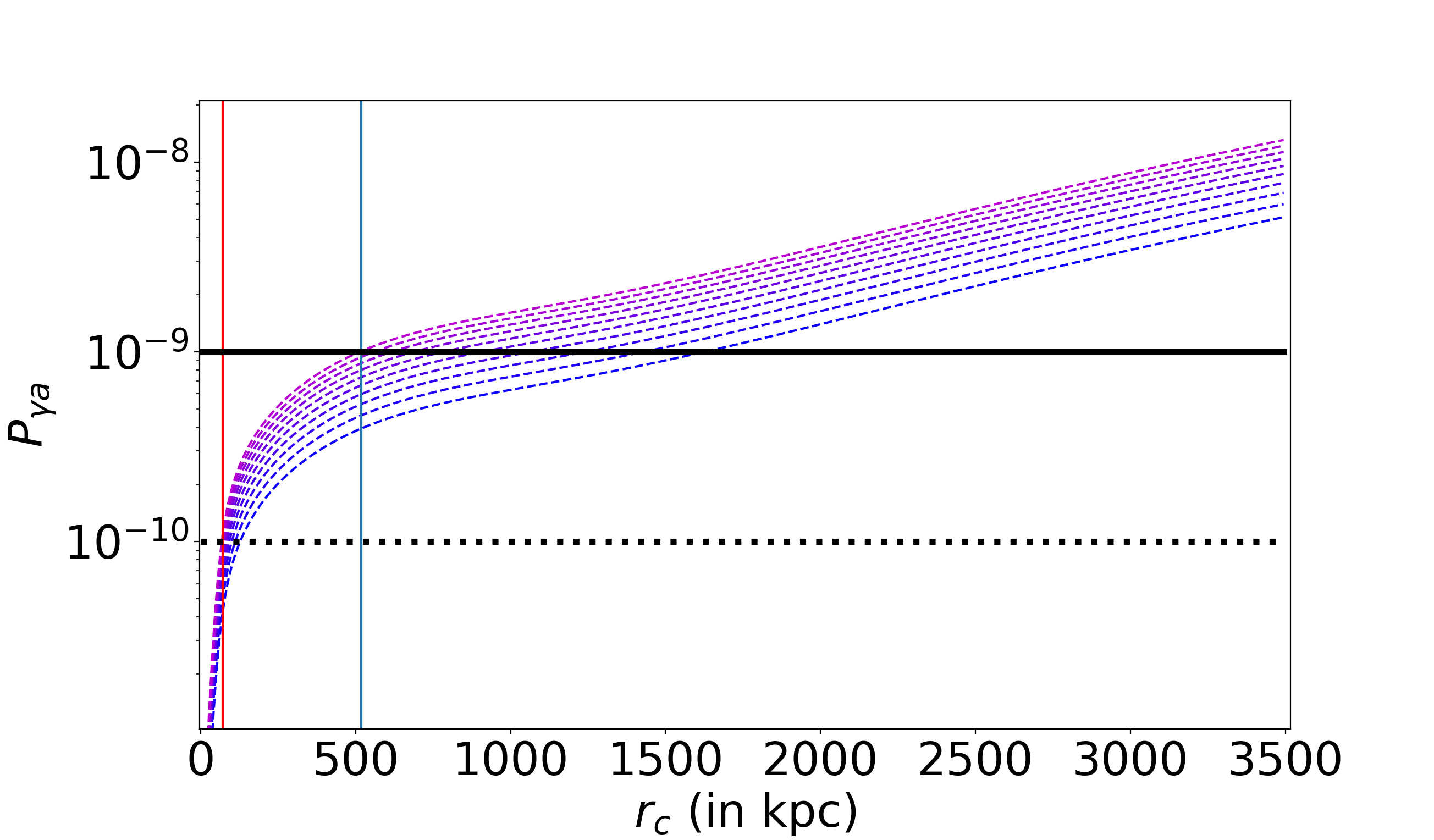}
     \caption{Signal strength for photon-ALP conversion for a range of redshifts $z$ from $0.1$ (bottom, blue) to $1.8$ (top, magenta) with $\Delta z=0.2$. Photons convert to ALPs at a radius corresponding to the ALP mass, see Fig.~\ref{Fig:mass-r-1}. The black solid and dotted line indicates the typical future CMB instrument noise level of the order $10^{-9}$K and $10^{-10}$K, respectively. The distortion from a single cluster can therefore probe ALPs from the region right to the cyan and red lines respectively for the distortion strengths $10^{-9}$K and $10^{-10}$K. The dominant contribution to the distortion is going to arise for low masses, when conversion occurs far from the core of the galaxy cluster.}\label{Fig:pa-1}
     \end{subfigure}
     
  \begin{subfigure}{1.\linewidth}
  \centering
     \includegraphics[trim={0cm 0 0cm 0cm}, clip, width=0.88\textwidth]{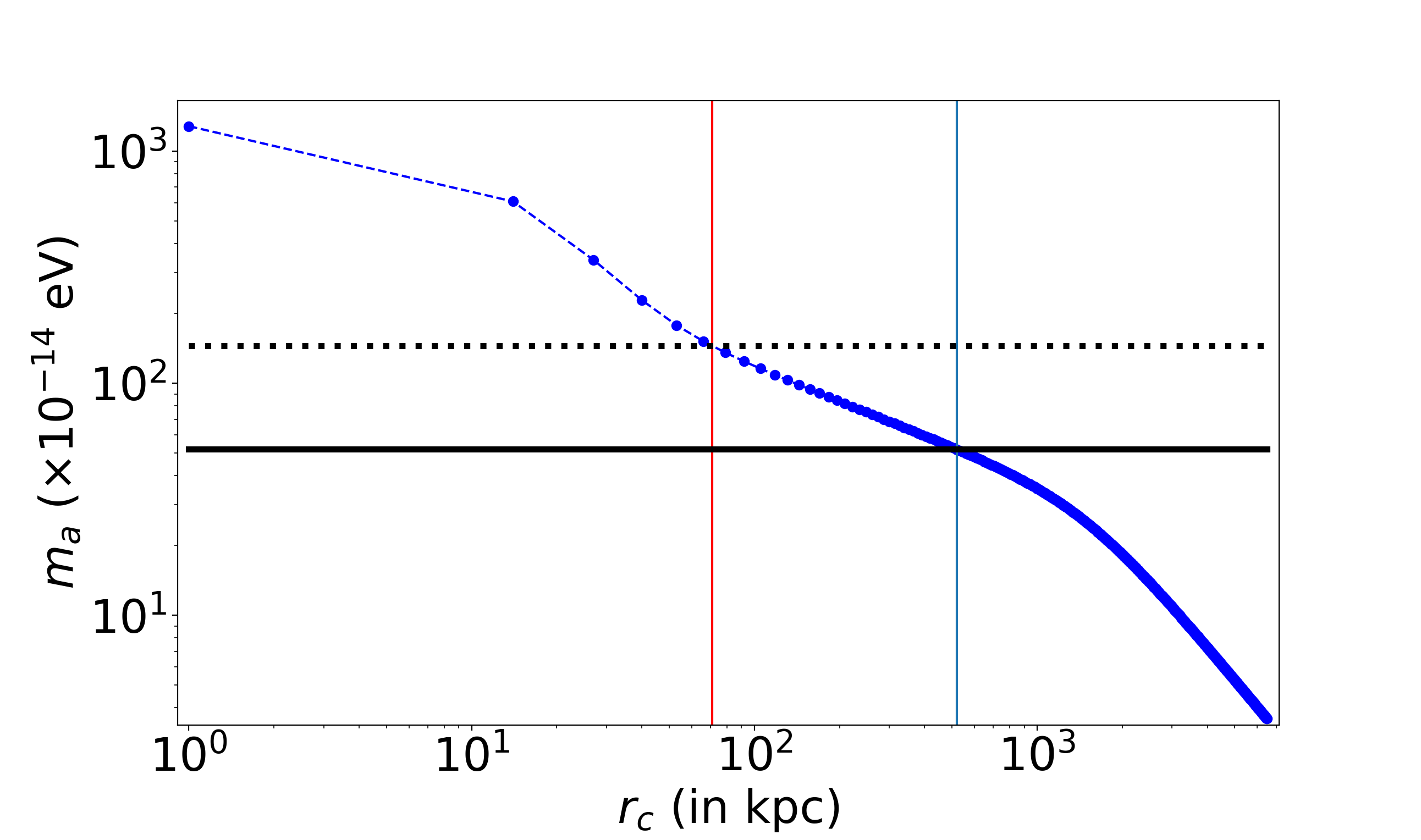}
     \caption{The mass of the ALPs and the radius at which the resonant conversion takes place are shown above for the model of electron density shown in Eq.~\eqref{elec-den}. The black solid and dotted line resembles the lines drawn in Fig.~\ref{Fig:pa-1}. The region below the solid and dotted lines produce distortions more than $10^{-9}$K and $10^{-10}$K respectively. The corresponding minimum radius of the galaxy cluster for which the resonant conversion can take place is shown by the cyan and red lines. This plot indicates that the lighter ALPs produce photon-ALPs distortion signal away from the core of galaxy clusters with a higher strength of the distortion (as can be seen from Fig.~\ref{Fig:pa-1}).}\label{Fig:mass-r-1}
     \end{subfigure}
     \caption{Resonant photon-ALP conversion signal as a function of radius of the galaxy cluster from its center are shown above for $g_{\gammaa}= 10^{-13}$ GeV$^{-1}$. }\label{Fig:pa-mass-alps}
\end{figure}

We solve Eq.~\eqref{net-prob-2} and Eq.~\eqref{Eq:gad-2} to calculate the photon-ALP conversion signal in the galaxy cluster scenario using electron density and magnetic field model mentioned in Sec.~\ref{ALP-cluster}. The estimated signal as a function of the cluster radius from its center is shown in Fig.~\ref{Fig:pa-1}. The variation of the signal with the redshift of the galaxy cluster (over the range $0.1-2.0$)  is also shown in Fig.~\ref{Fig:pa-1}. A future CMB experiment with a noise level of $10^{-9}$K, can probe $P_{\gammaa}$ of the same order,\footnote{assuming no foreground contamination} which can be translated into a minimum radius $r_{\text{min}}$ and hence a maximum axion mass $(m_a)_{max}$. This is shown by the black solid line and the cyan line in Fig.~\ref{Fig:pa-1}. The regions on the right side of the cyan line and above the black solid line are accessible to measure from galaxy clusters with the next generation CMB experiments having instrument noise $10^{-9}$K. However, the spectral distortion signal of the order $10^{-10}$ is also possible to reach technologically and in that case, we can probe to a radius of $70$kpc, which is shown by the black dotted line and the corresponding radius is shown by the red line.
From Eq.~\eqref{net-prob-2} and Eq.~\eqref{Eq:gad-2}, we see that the distortion is proportional to the magnetic field strength and is inversely proportional to the electron density. As a result, for the low-mass ALPs the resonant conversion takes place far from the core of the cluster resulting into a stronger conversion of the signal in contrast to the case of heavier ALPs. 

\begin{figure}[H]
\centering
    \includegraphics[trim={0 0 0 0cm}, clip, width=1\linewidth]{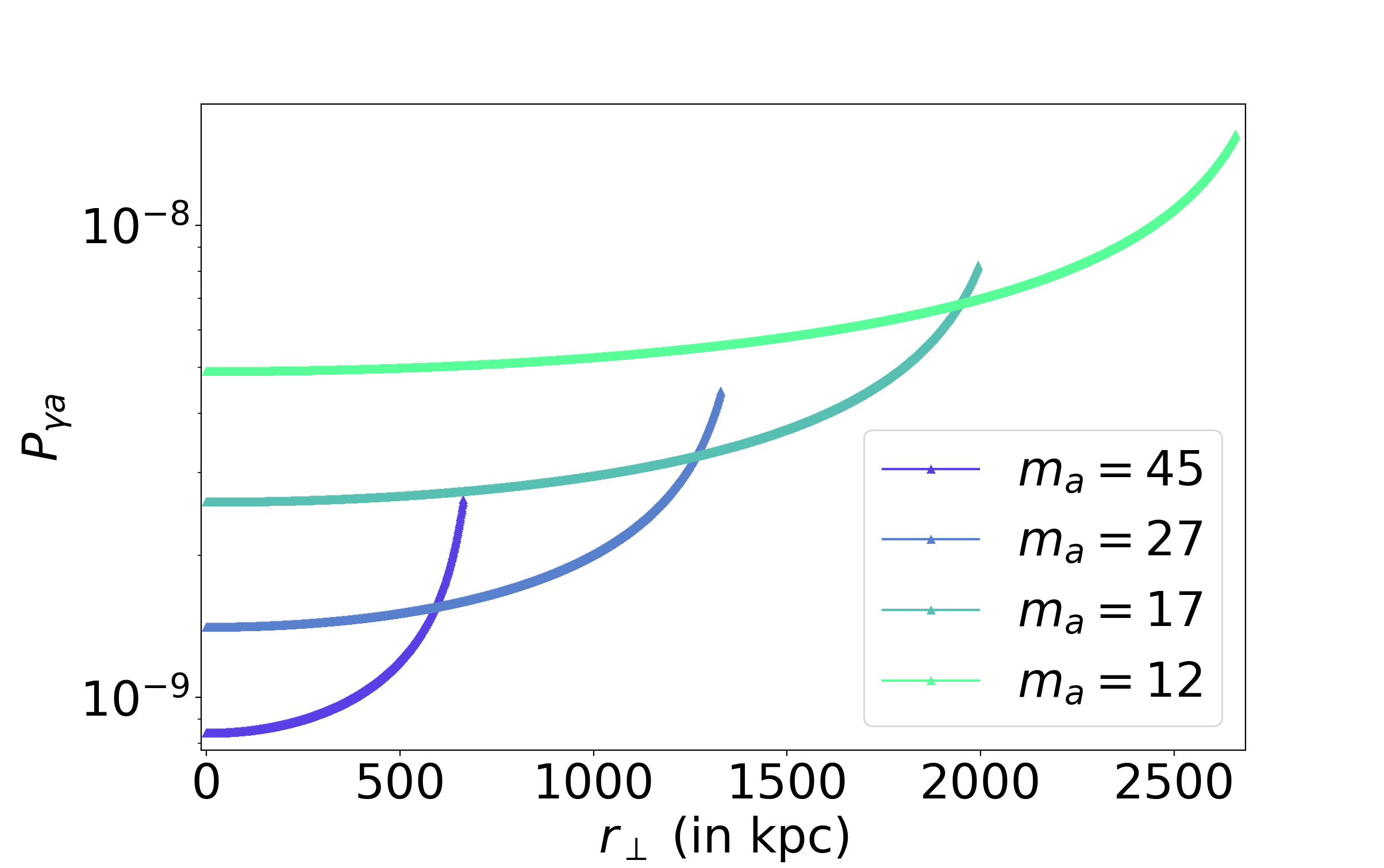}
     \captionsetup{singlelinecheck=on,justification=raggedright}
 \caption{Radial signal profiles $P_\gammaa$ of the projected conversion spheres (see Fig.~\ref{Fig:mass-shell}) around the clusters for four ALP masses ($m_a$ in units of $10^{-14}$ eV) as a function of the radius the cluster center perpendicular to the line of sight. We have used the value of $g_{\gammaa}= 10^{-13}$ GeV$^{-1}$ to obtain this plot.}\label{Fig:pgamma_x}
\end{figure}
The ALP signal at different radius originates from different ALP masses for which $m_\gamma= m_a$. In Fig.~\ref{Fig:mass-r-1}, we plot the values of ALP mass as a function of radial distance from the center of the galaxy cluster at which we expect a resonant conversion to take place. By comparing Fig.~\ref{Fig:pa-1} and Fig.~\ref{Fig:mass-r-1}, we see that the range of radius for which the distortion $P_{\gamma a}\geq 10^{-9}$, corresponds to a mass range below $5\times 10^{-13}$ eV (the region below the black line and to the right side of the cyan-line). 

\begin{figure}
\begin{subfigure}{0.5\linewidth}
  \centering
     \includegraphics[trim={0cm 0 0cm 0cm}, clip, width=1.\textwidth]{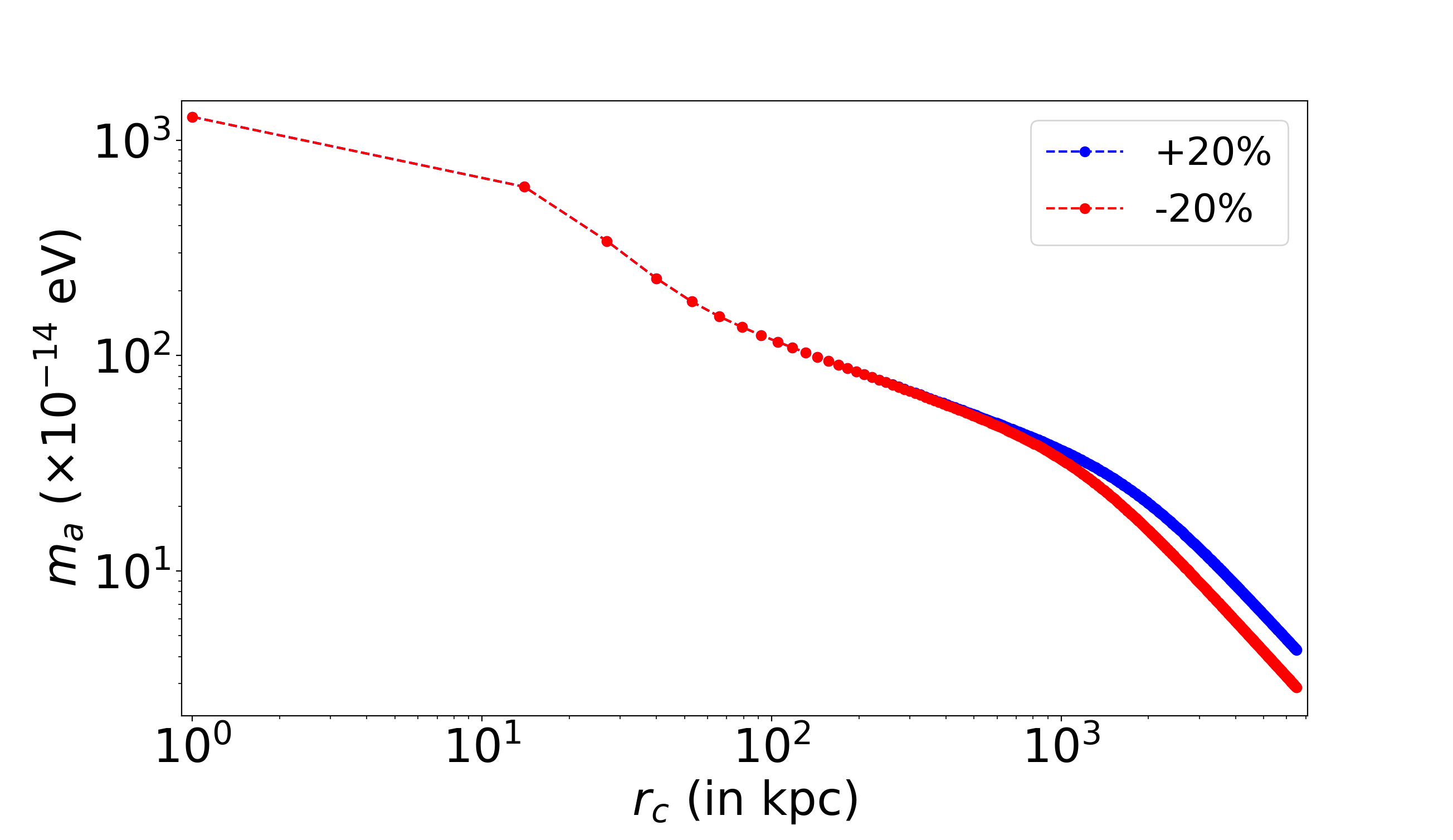}
     \caption{Variation in $r_s$}\label{Fig:mass-rs-1}
     \end{subfigure}
     \begin{subfigure}{0.5\linewidth}
  \centering
     \includegraphics[trim={0cm 0 0cm 0cm}, clip, width=1.\textwidth]{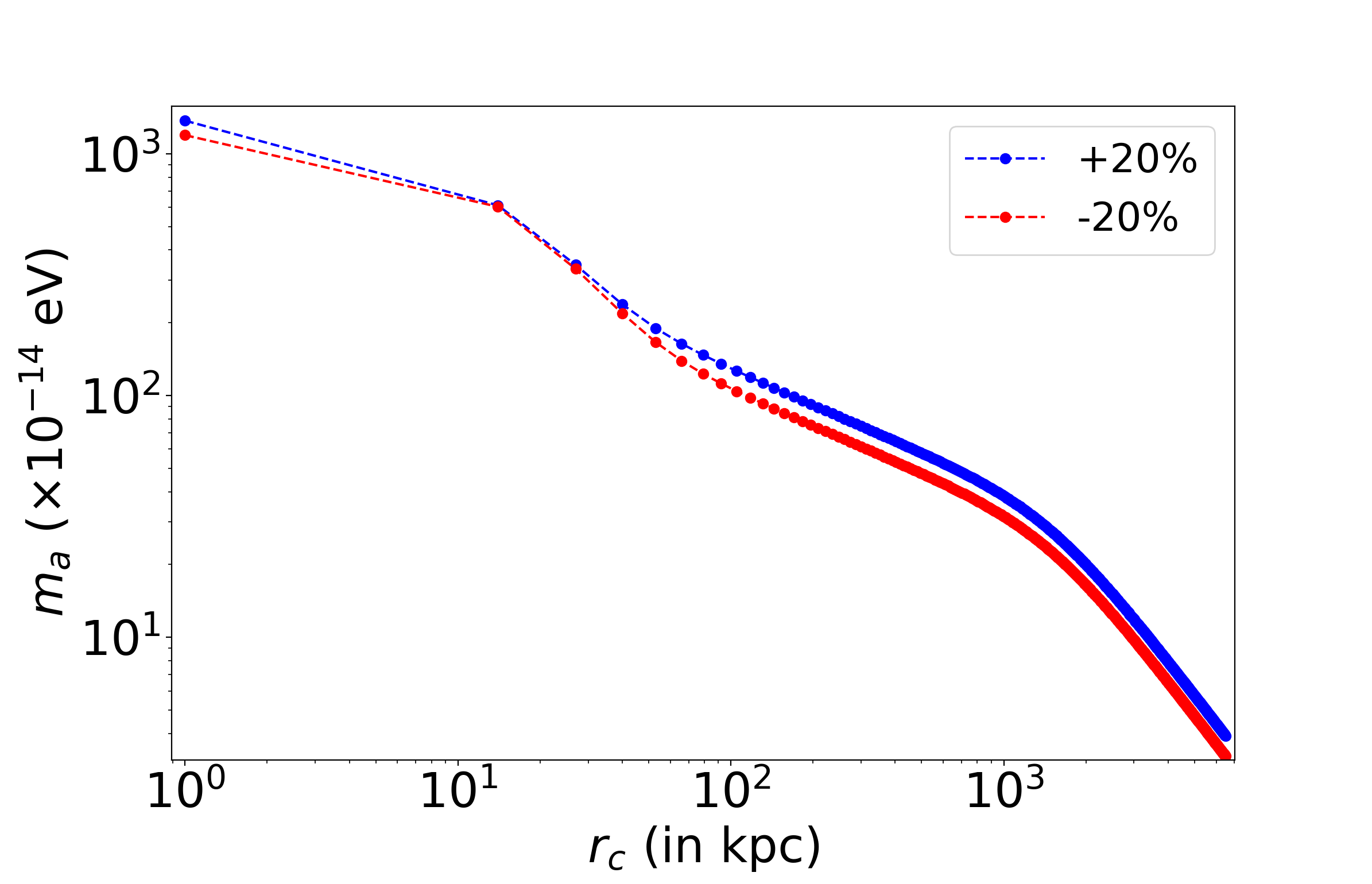}
     \caption{Variation in $r_{c1}$}\label{Fig:mass-rc-1}
     \end{subfigure}
     \begin{subfigure}{0.5\linewidth}
  \centering
     \includegraphics[trim={0cm 0 0cm 0cm}, clip, width=1.\textwidth]{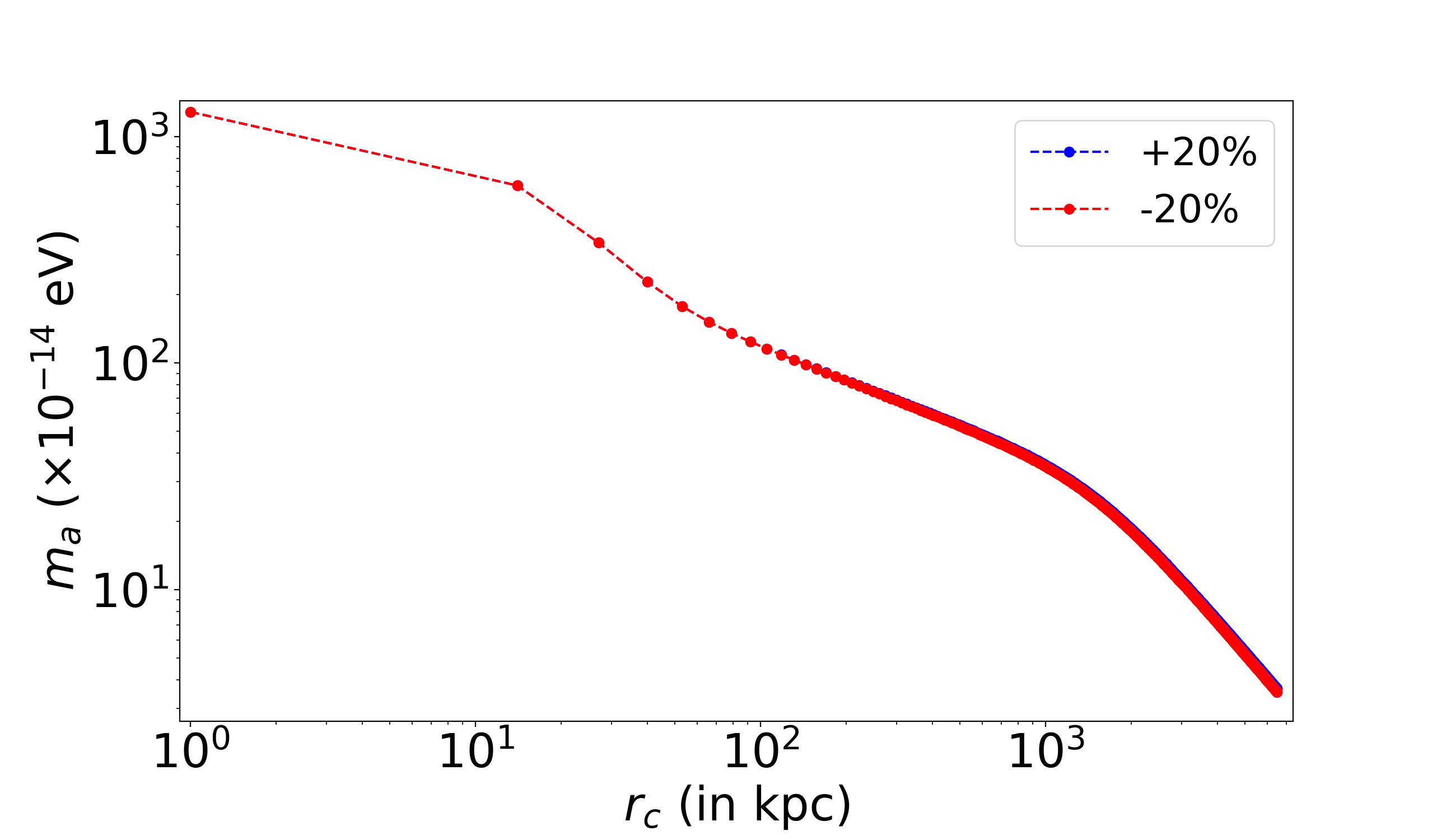}
     \caption{Variation in $\beta$}\label{Fig:mass-beta-1}
     \end{subfigure}
     \begin{subfigure}{0.5\linewidth}
  \centering
     \includegraphics[trim={0cm 0 0cm 0cm}, clip, width=1.\textwidth]{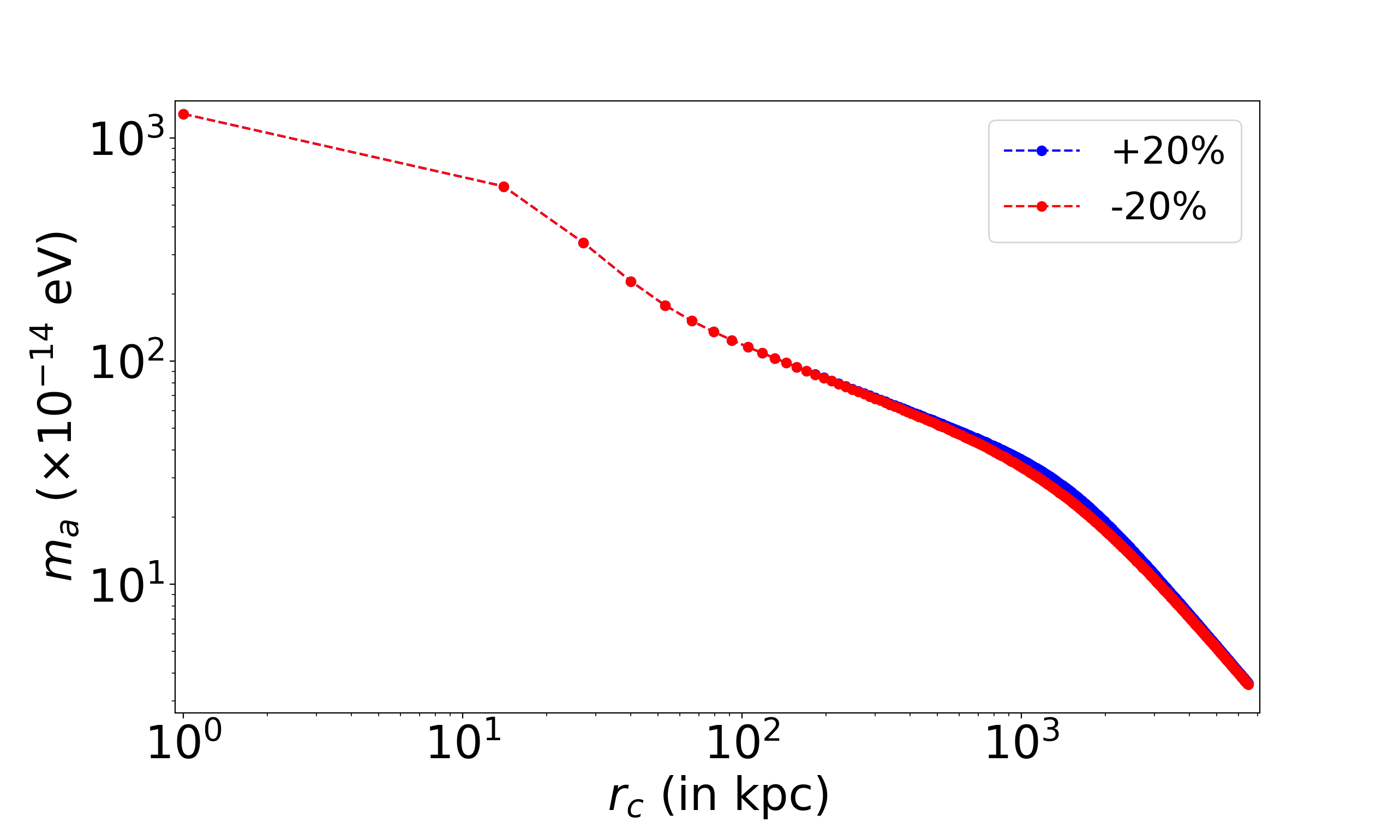}
     \caption{Variation in $\gamma$}\label{Fig:mass-gam-1}
     \end{subfigure}
     \begin{subfigure}{0.5\linewidth}
  \centering
     \includegraphics[trim={0cm 0 0cm 0cm}, clip, width=1.\textwidth]{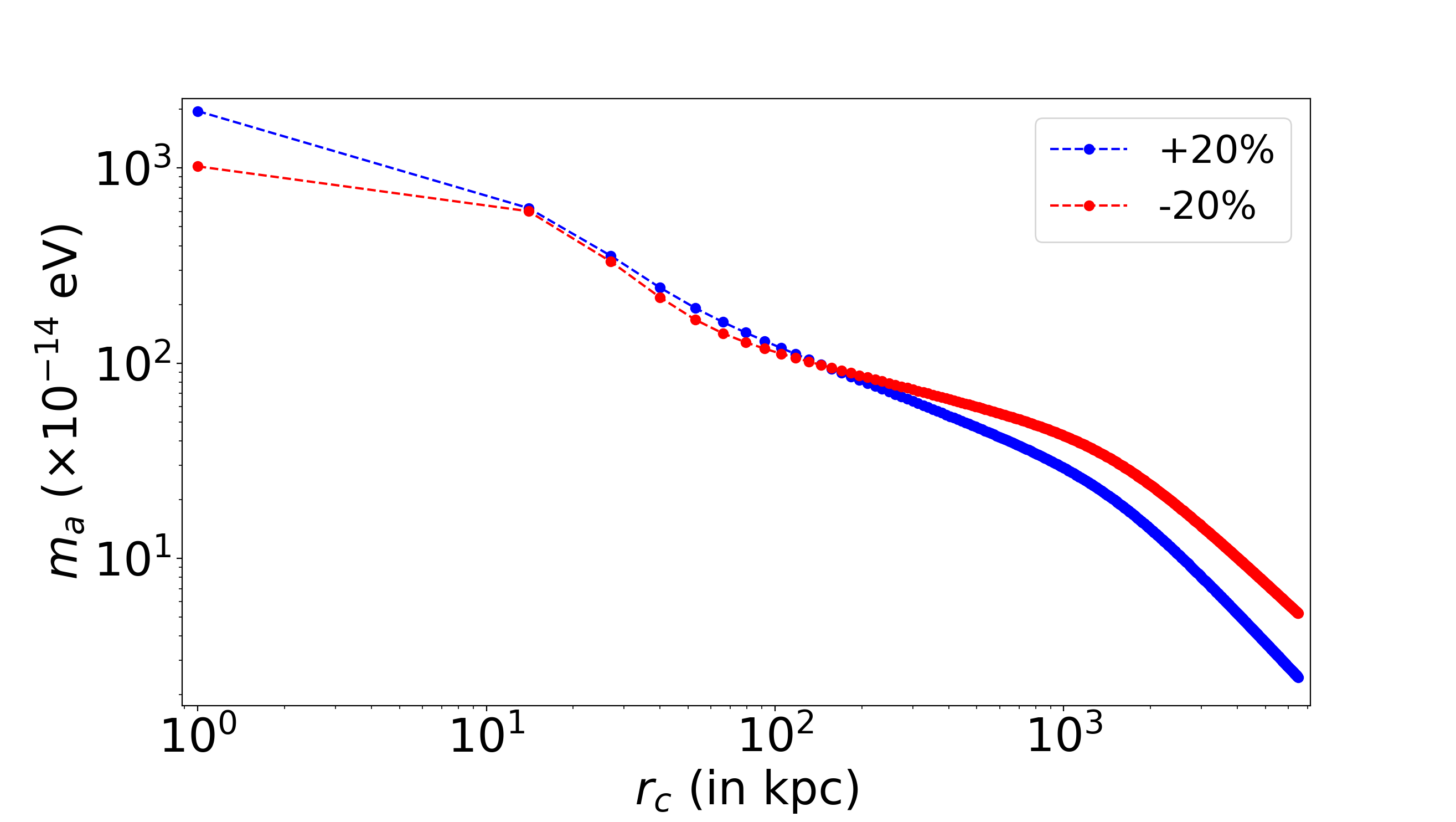}
     \caption{Variation in $\alpha$}\label{Fig:mass-alpha-1}
     \end{subfigure}
     \begin{subfigure}{0.5\linewidth}
  \centering
     \includegraphics[trim={0cm 0 0cm 0cm}, clip, width=1.\textwidth]{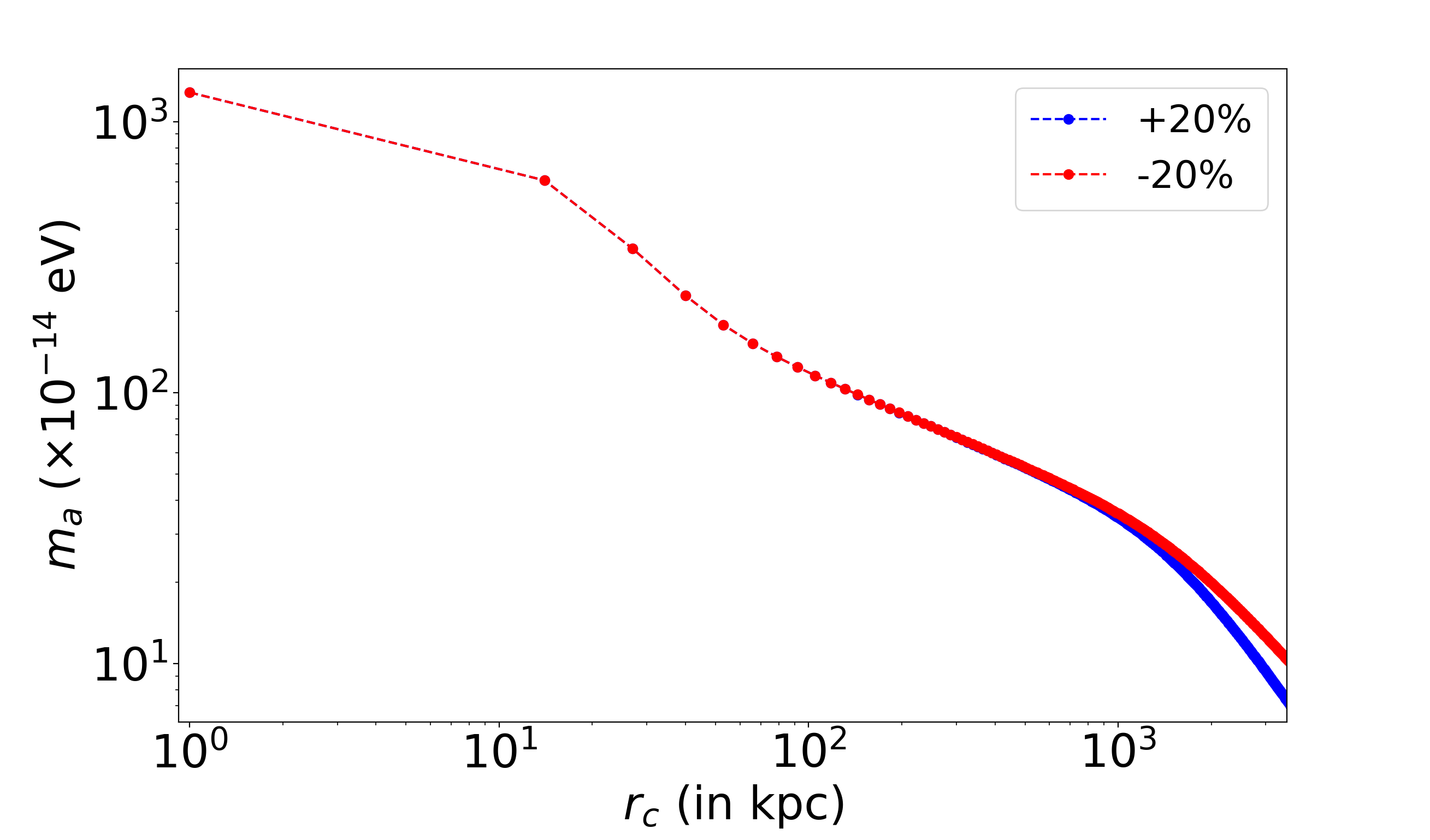}
     \caption{Variation in $\epsilon$}\label{Fig:mass-alpha-1}
     \end{subfigure}
     \caption{ {Variation in the radius at which the resonant photon-ALP conversion takes place to a particular ALP mass $m_a$ due to the variations in the parameters of the model of electron density are shown by changing the parameters $r_s, r_{c1}, \beta, \gamma, \alpha, \epsilon$ by $\pm 20\%$ from the fiducial values $r_s= 1000$ kpc, $r_{c1}= 100$ kpc, $\beta=0.64$, $\gamma=3$, $\alpha=2$ and $\epsilon=4$.}}\label{Fig:elec_var}
\end{figure}

The spatial profile of $P_{\gammaa}$ as a function of the radius perpendicular to the line of sight (i.e. cluster radius projected on the sky) is shown for four ALP masses $m_a (\times 10^{-14} eV)= 45,\, 27,\, 17,\, 12$ is shown in Fig.~\ref{Fig:pgamma_x}. ALPs with higher mass gets converted close to the cluster center with a lower value of $P_\gammaa$ and has smaller angular size on the sky. As a result, the signal projected on the sky will appear like a disc of different radius depending upon the ALP mass. 

\begin{figure}
    \centering
    \includegraphics[trim={0 0 0 0cm}, clip, width=1\linewidth]{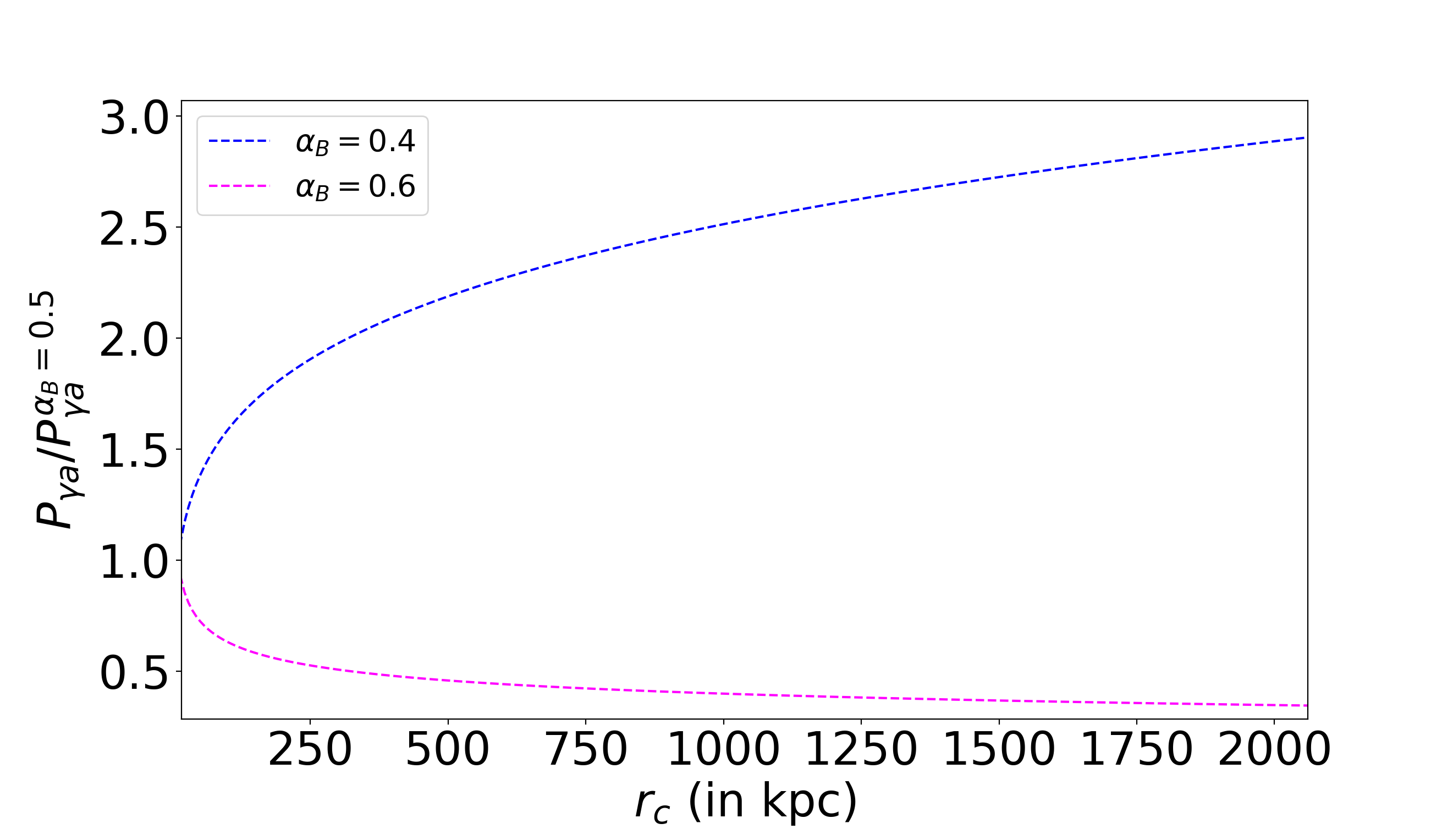}
    \caption{ {We show the variation in the value of $P_{\gamma a}$ for the variation in the radial dependence of the magnetic field value $B(r)= B_0(r/r_b)^{-\alpha_B}.$}}
    \label{fig:mag_var}
\end{figure}

 {The variation of the electron density in galaxy clusters can lead to variation in the signal strength and shape of the distortion. We show in Fig. \ref{Fig:elec_var} the variation in the radius at which the resonant conversion takes place to a ALP mass $m_a$ by varying the electron density parameters such as $r_c$, $r_s$, $\alpha$, $\beta$, $\gamma$ and $\epsilon$. These parameters typically affect the values for the ALPs masses in the range $10^{-12}-10^{-13}$ eV, which is the interested regime for this analysis \footnote{See Fig. \ref{Fig:mass-g-snr} for the interested ALP mass regime for the upcoming CMB missions.} The variation of the magnetic field in the galaxy clusters can also lead to variation in the photon-ALPs conversion probability $P_{\gamma a}$. We show the variation of $P_{\gamma a}$ for different values of the magnetic field radial dependence $B(r)= B_0(r/r_b)^{-\alpha_B}$ in Fig. \ref{fig:mag_var}. A variation in the value of $\alpha_B$ by $\pm 20\%$ leads to about a factor of three variation in the value of $P_{\gamma a}$ for $r_c$ up to $2000$ kpc. Change in the value of $B_0$ and $r_b$ leads to only a constant shift in the value of $P_{\gamma a}$.} 

The spectral shape of the photon-ALP distortion signal also carries a  spectral shape different from other polarized foreground contamination like synchrotron and dust. The frequency spectrum of the photon-ALP conversion signal along with those of potential contaminants is depicted in Fig.~\ref{Fig:Inu-nu} for CMB-S4 \cite{Abazajian:2016yjj, 2019arXiv190704473A}  frequency channels. The shape of the distortion resembles  the primary CMB anisotropy which is therefore a potential contaminant of the photon-ALP distortion signal.

Along with this signal at the location of the clusters, other spectral distortion and secondary anisotropies are generated around clusters such as  tSZ, kSZ, and relativistic-SZ. These distortions differ from the photon-ALP distortion signal in multiple ways: (i) the shape of the spectral distortion signal is different; (ii) 
spectral distortion signal from the ALP is $100\%$ polarized and remains highly polarized even in the case of multiple resonances; (iii) the photon-ALP conversion gives a characteristic radial profile that grows with distance from the cluster center and then drops to zero.

One of the potential source of contamination is due to scattering of the local CMB quadrupole at the location of the cluster by the electrons in the galaxy cluster \cite{Hall:2014wna,Louis:2017hoh}, and hence depends upon the optical depth $\tau (\hat n)$ in the galaxy clusters. Using the model of electron density mentioned in Sec.~\ref{ALP-cluster}, we estimate the radial profile of $\tau (\hat n)$ \cite{Hall:2014wna,Louis:2017hoh}
\begin{equation}
\tau (\theta, z)= \int \sigma_T n_e(r(l), z) dl \approx 10^{-3}- 10^{-4}, \label{optical_depth_cl}
\end{equation}
where $r(l)= \sqrt{l^2 + (D_A\theta)^2}$ and  $\sigma_T$ is the Thompson scattering cross-section.  For a local CMB temperature quadrupole  fluctuation $\sqrt{C_2} \approx 20 \, \mu$K, we can expect a RMS polarization fluctuation of the order $\mathcal{P}_Q \approx 2\tau \mu$K. Cluster have $\tau \approx 10^{-3}-10^{-4}$ giving the polarization contamination $10^{-3}-10^{-4}\, \mu$K. For massive clusters, contamination from the scattering of CMB quadrupole will become important and may limit the sensitivity of photon-ALP conversion. One way to push the sensitivity when we hit this limit would be to use only the low optical depth clusters. Another approach will be to use the correlation of  $\mathcal{P}_Q$ with the CMB quadrupole at the source redshift of the cluster \cite{Hall:2014wna,Louis:2017hoh} to separate the axion distortion from $\mathcal{P}_Q$.
In the next section, we will discuss the possible contamination from several foregrounds and will study the realistic appearance of the photon-ALP distortion signal in the simulated sky.

\section{Simulated sky map of photon-ALPs distortion in the presence of galactic foregrounds, CMB anisotropy and instrument noise}\label{ALP-contamination}
The microwave sky contains several astrophysical signals originating from galactic and extragalactic sources and is a contamination to the spectral distortion signals \cite{2016A&A...594A..10P, Thorne:2016ifb, Abitbol:2017vwa}. These astrophysical signals like synchrotron emission, thermal dust, Anomalous Magnetic Emission (AME), free-free, etc. are dominant in different frequency ranges of the microwave sky \cite{Abitbol:2017vwa}. To measure the spatially fluctuating part of spectral distortions of CMB, we also need to remove primordial CMB, along with tSZ and kSZ. In this section, we explore the foreground contamination to the photon-APLs distortion signal and simulate realistic maps of the photon-ALP distortion in presence of these foregrounds.

\subsection{Simulation of the photon-ALPs distortion signal}
The photon-ALPs spectral distortion signal around every cluster given in Eq.~\eqref{Eq:bb-dist} depends on the electron density and the magnetic field of the galaxy cluster as discussed in Sec.~\ref{amp_alps}. In this analysis, we have assumed an universal profile of the cluster electron density as discussed in Sec.~\ref{elec-den} for a typical cluster of mass $\sim 10^{14}\, M_\odot$ which will be probed by the upcoming missions such as SO  \cite{Ade:2018sbj} and CMB-S4 \cite{Abazajian:2016yjj, 2019arXiv190704473A} .  The strength of magnetic field in the galaxy cluster is taken as an universal profile as discussed in Sec.~\ref{mag-field} and the direction of the magnetic field is considered to be randomly oriented. As a result, the polarization direction of the ALP distortion (which is parallel to the magnetic field direction) is also going to be randomly oriented. But the strength of the distortion which depends on the magnitude of the magnetic field $B^2$ is going to follow the radial profile as given in Eq. \ref{mag-field}. The spatial shape of the ALP signal for a fixed ALP mass around every cluster localization is a disc with varying sizes due to different angular diameter distance to the cluster. The strength of the ALPs distortion signal and the size of the disc increases with decrease in the mass of the ALPs, as shown in Fig.~\ref{Fig:pgamma_x}.
We have taken the position of the galaxy clusters distributed randomly on the sky outside the galactic plane with the redshift distribution following the $N(z)/\Delta z$ as shown in Fig.~\ref{Fig:dn-dz}. 

\subsection{Simulations of galactic foregrounds and CMB primary and secondary anisotropies}
\begin{figure}[h]
\centering
\begin{subfigure}{1\linewidth}
\includegraphics[trim={0 0 0cm 0.8cm}, clip, width=0.5\textwidth]{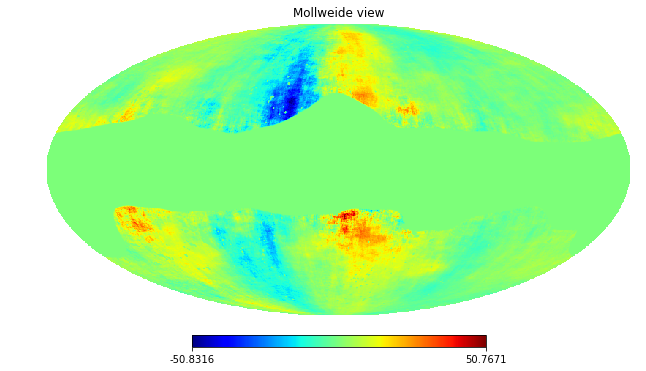}
\includegraphics[trim={0 0 0cm 0.8cm}, clip, width=0.5\textwidth]{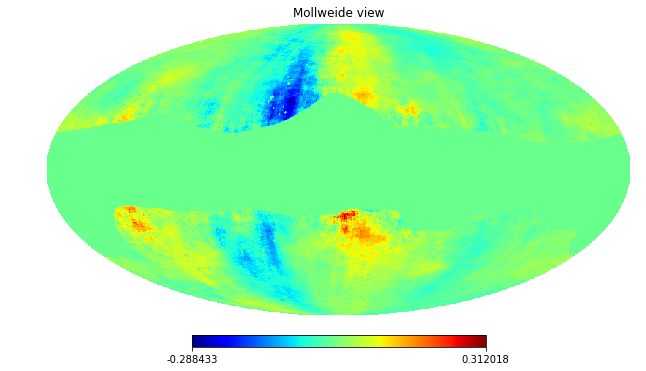}
     \caption{Synchrotron polarization $Q$ map at $\nu=28$ GHz and $\nu=150$ GHz from PySM.}\label{Fig:foreground-map-1}
     \end{subfigure}
  \begin{subfigure}{1\linewidth}
     \includegraphics[trim={0 0 0 0.8cm}, clip, width=0.5\textwidth]{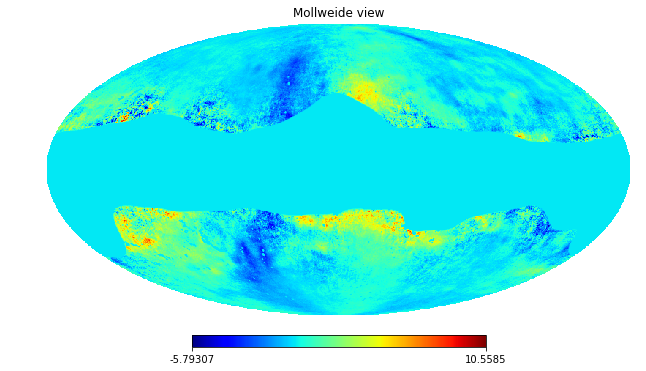}
          \includegraphics[trim={0 0 0 0.8cm}, clip, width=0.5\textwidth]{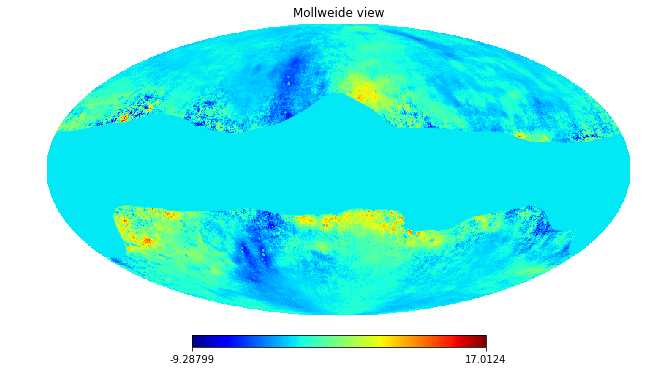}
     \caption{Dust polarization $Q$ map at  $\nu=150$ GHz and $\nu=230$ GHz  from PySM.}\label{Fig:foreground-map-2}
     \end{subfigure}
  \caption{$30\%$ sky fraction is masked to remove the contamination from the galactic plane.}\label{Fig:foreground-map}
\end{figure}

The primary sources of contamination to the polarization signal of photon-ALPs conversion is due to synchrotron, dust and CMB anisotropies. TSZ and kSZ signals are not polarized, and we ignore there contribution in this analysis.
We use the publicly available foreground simulation code Python Sky model (PySM) \cite{Thorne:2016ifb}  to estimate the foregrounds sky with the Model s-3 parameters for synchrotron power-law with a curved index \cite{Thorne:2016ifb} and, we use Model d-5 for dust \cite{Thorne:2016ifb}. The c-1 model is used to generate the CMB sky maps \cite{Thorne:2016ifb, pysm}. Details for the code and models are provided on the PySM website \cite{pysm}. We consider the polarization map (Stokes $Q$ and $U$) 
for synchrotron and thermal dust from PySM at the SO \cite{Ade:2018sbj} and CMB-S4 \cite{Abazajian:2016yjj, 2019arXiv190704473A}  frequency channels (mentioned in Sec.~\ref{s4noise}). The corresponding $Q$ and $U$ maps (in Rayleigh-Jeans temperature $K_{RJ}$ units) for synchrotron and dust at a few frequency channels are shown in Fig.~\ref{Fig:foreground-map}. We masked the galactic plane covering the $30\%$ sky fraction to reduce the foreground contamination. The polarised SZ effect \cite{1999MNRAS.310..765S, Louis:2017hoh} can also be a contamination for this signal but will be weaker than other contamination like CMB, synchrotron and dust and we ignore the effect from polarised SZ in this forecast. 

\subsection{Instrument noise for Simons Observatory and CMB-S4 }\label{s4noise}
Simons Observatory (SO) \cite{Ade:2018sbj} is an upcoming ground-based CMB experiment located at Cerro Toco in Chile, going to be operational from the early 2020s. The science goals of SO \cite{Ade:2018sbj} encompasses a wide range of topics and a detailed study of this can be found in  \cite{Ade:2018sbj}. SO \cite{Ade:2018sbj} has six frequency channels covering a range from $27-280$ GHz. The goal-instrument noise and beam resolution for the Large Aperture Telescope (LAT) setup are shown in Table \ref{tab:so}. 

CMB-S4 \cite{Abazajian:2016yjj, 2019arXiv190704473A} is a ground-based CMB mission proposed to measure the microwave sky with about $10^5$ high sensitivity detectors and angular resolution of typically one arcminute. The desired detector noise specifications and frequency channels are shown in Table \ref{tab:s4}.  

  \begin{figure}[h]
     \includegraphics[trim={0 0 0 0cm}, clip, width=1\textwidth]{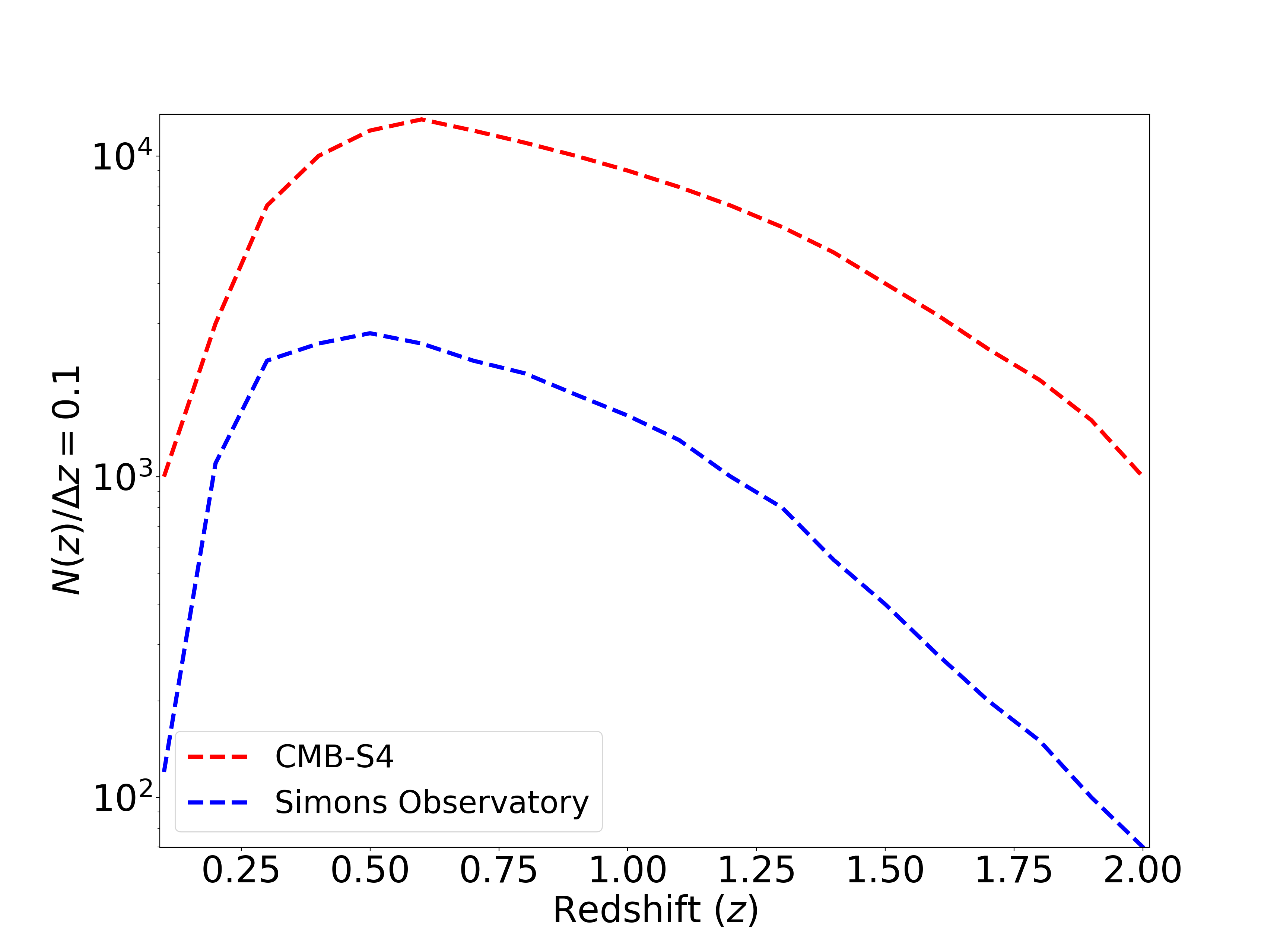}
          \captionsetup{singlelinecheck=on,justification=raggedright}
     \caption{The number distribution of galaxy clusters as a function of redshift assumed in this analysis for Simons Observatory \cite{Ade:2018sbj} and CMB-S4 \cite{Abazajian:2016yjj, 2019arXiv190704473A}  are shown above. }\label{Fig:dn-dz}
     \end{figure}

\begin{table}[H]
\centering
\captionof{table}{Simons Observatory noise used in the analysis with  $\Delta P= \sqrt{2}\Delta T$ \cite{Ade:2018sbj}. We have used only $40\%$ sky in this analysis.} \label{tab:so}

\begin{tabular}{ |m{2.0cm}|m{2.cm}|m{2.0cm}|}
\hline
 $\nu$ (GHz) & $\Delta T$ ($\mu\, K_{CMB}$-arcmin) & Beam (arcmin)\\
 \hline
27  & 52 & $7.4$ \\
\hline
39  & 27 & $5.1$ \\
\hline
93  & 5.8 & $2.2$ \\
\hline
145  & 6.3 & $1.4$ \\
\hline
225  & 15 & $1.0$ \\
\hline
280  & 37 & $0.9$ \\
\hline
\end{tabular}
\end{table}

\begin{table}[H]
\centering
\captionof{table}{CMB-S4 \cite{Abazajian:2016yjj, 2019arXiv190704473A} noise used in the analysis with  $\Delta P= \sqrt{2}\Delta T$. The exact value of the CMB-S4 noise is going to vary from the current values depending upon the real design. We have used only $70\%$ sky in this analysis.} \label{tab:s4}
\begin{tabular}{ |m{2.0cm}|m{2.cm}|m{2.0cm}|}
\hline
 $\nu$ (GHz) & $\Delta T$ ($\mu\, K_{CMB}$-arcmin) & Beam (arcmin)\\
 \hline
28  & 20 & $5.1$ \\
\hline
41  & 17.5 & $3.4$ \\
\hline
90  & 2 & $1.6$ \\
\hline
150  & 1.8 & $1.0$ \\
\hline
230  & 6.3 & $0.6$ \\
\hline
\end{tabular}
\end{table}

\begin{figure}[H]
\centering
\begin{subfigure}{1\textwidth}
\centering
\includegraphics[trim={0 0 0 0cm}, clip, width=0.9\textwidth]{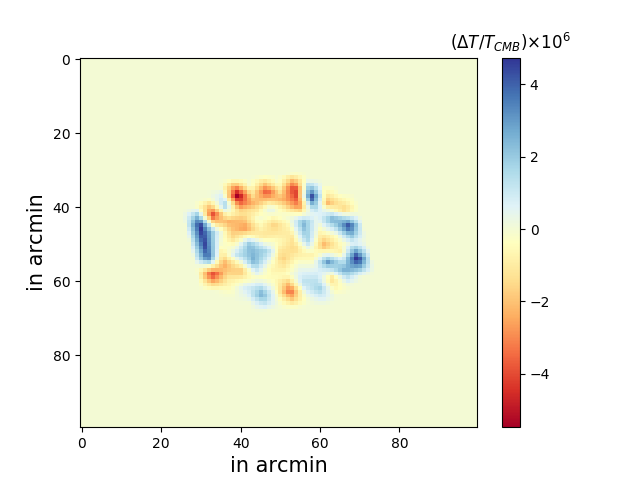}
\caption{The photon-ALP distortion signal without any foreground contamination and instrument noise. The spatial  anisotropy in the photon-ALPs distortion arises due to the spatial variation in the electron density and magnetic field. We have taken the spatial profile of the electron density and magnetic field as described in Eq. \ref{elec-den} and Eq. \ref{mag-field} respectively. The direction of the magnetic field is assumed to be randomly oriented at every location, a conservative assumption because it will maximally reduce the observable signal after averaging with the beam. }\label{Fig:ALP-map-1-pure}
\end{subfigure}
\begin{subfigure}{1\textwidth}
\centering
\includegraphics[trim={0 0 0 0cm}, clip, width=0.9\textwidth]{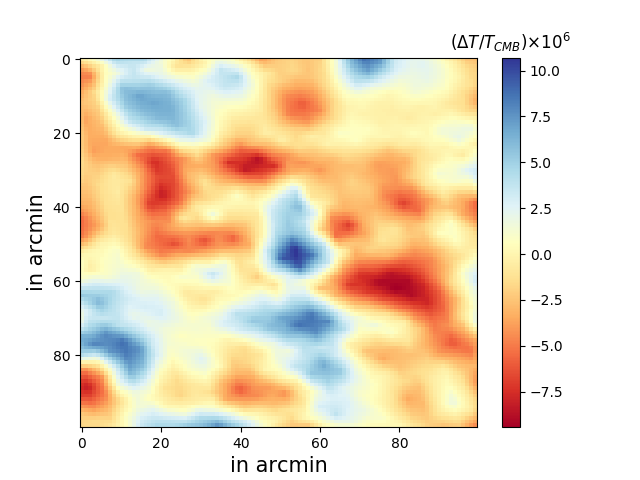}
\caption{Sky signal at the frequency channel $\nu= 230$ GHz in the presence of foreground contamination and instrument noise of CMB-S4 \cite{Abazajian:2016yjj, 2019arXiv190704473A} given in table \ref{tab:s4}.}\label{Fig:ALP-map-1-impure}
\end{subfigure}
\caption{The $Q$ map of the photon-ALP distortion for the axion mass $m_a= 10^{-13}$ eV and photon-ALP coupling $g_{\gammaa}= 10^{-12}$  GeV$^{-1}$ for a single cluster at redshift $z=0.2$. }\label{Fig:ALP-map-1a}
\end{figure}

The high-frequency channels of both these missions are expected to be useful for cleaning the dust, whereas the low-frequency channels are going to be useful for cleaning the synchrotron. In order to capture the effect from the instrument noise, we add Gaussian instrument noise using the specifications provided in Table \ref{tab:so} and \ref{tab:s4}. As the PySM software \cite{Thorne:2016ifb, pysm} requires the instrument noise in the Rayleigh-Jeans temperature units ($T_{RJ}$), we convert the instrument noise provided in Table \ref{tab:so} and \ref{tab:s4} into $T_{RJ}$ by using the relation 
\begin{equation}\label{tcmb-trj}
\Delta T_{RJ}= \Delta T_{CMB}\frac{f^2e^f}{(e^f-1)^2},
\end{equation}
where $f= h\nu/k_BT_{CMB}$ and $T_{CMB}= 2.7255$K and $h,\,k_B$ are the Planck's constant and Boltzmann constant respectively. 
In the next section, we discuss the ALP signal in presence of these contamination and simulate a realistic sky map.
\subsection{Sky model in the presence of photon-ALP distortion, foregrounds and instrument noise}
The photon-ALP distortion signal arising from galaxy clusters is going to be highly distorted from the actual predicted shape due to the strong contamination from foregrounds and instrument noise. A realistic sky signal after accounting for the foregrounds and instrument noise can be written as \footnote{Bold notations denote matrix.}
\begin{align}\label{ilc-1}
\begin{split}
S^{\{Q, U\}}_{\nu_i}(\hat p)&= A_{\nu_i j}x^{\{Q, U\}}_{j}(\hat p) + n_{\nu_i} (\hat p),\\
\mathbf{S}^{\{Q, U\}} (\hat p)&= \mathbf{A}\mathbf{x}^{\{Q, U\}}(\hat p) + \mathbf{n} (\hat p),
\end{split}
\end{align}
where $\mathbf{S}^{\{Q, U\}}(\hat p)$ is the polarization map $Q (\hat p)$ and $U (\hat p)$ in the sky direction $\hat p$, $A_{\nu_i j}$ is the mixing matrix and $x^{\{Q, U\}}_{j}$ are different components with $j \in$ \{CMB, ALP, synchrotron, dust\}. Following Eq.~\eqref{ilc-1}, we add simulated ALP signal at a randomly selected sky localization outside the  galactic plane (which are assumed to be the cluster locations) along with CMB, foreground and instrument noise.  

We show the maps of the ALP signal (for axion mass $10^{-13}$ eV and $g_{\gamma a}= 10^{-12}$ GeV$^{-1}$) around a single cluster for two cases, (i) only the sky signal of photon-ALP distortion without any contamination (Fig.~\ref{Fig:ALP-map-1-pure}), (ii) photon-ALP distortion signal in the presence of foreground contamination and instrument noise (Fig.~\ref{Fig:ALP-map-1-impure}). The cluster location is assumed to be at redshift $z=0.2$. The size of the disc of the signal depends on the radius at which the resonant conversion is taking place and also on the angular diameter distance to the galaxy cluster. So, if the electron density of all the galaxy clusters has a universal spatial structure, then it will produce a same angular size at a fixed redshift. The electron density and the magnetic field are considered as described in Sec. \ref{ALP-cluster}. The directions of the magnetic field are considered to be randomly oriented in the galaxy cluster.  {Along each line of sight, the photon-ALPs conversion happens twice and the magnetic field direction is considered to be randomly oriented at both the conversions and varies for each line of sight. This is the source of the fluctuations in Fig. \ref{Fig:ALP-map-1-pure}, even in the absence of any contaminations.}  This is a conservative choice, and the polarization signal can be more prominent if the magnetic field is ordered within the instrument beam. Due to the presence of foreground contamination, the ALPs distortion is noise dominated as shown in Fig.~\ref{Fig:ALP-map-1-impure} and the presence of the signal is nearly impossible to detect for a single object without foreground cleaning.

\subsection{Cleaning the foregrounds using the  internal linear combination method}\label{ilc}
\begin{figure}[H]
\centering
\includegraphics[trim={0.5cm 0.2cm 0.5cm 0.5cm}, clip, width=.9\textwidth]{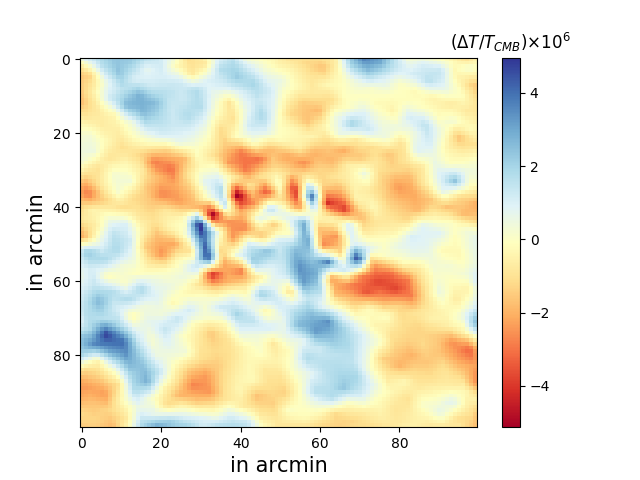}
\caption{The $Q$ map of the photon-ALP distortion for the axion mass $m_a= 10^{-13}$ eV and photon-ALP coupling $g_{\gammaa}= 10^{-12}$ GeV$^{-1}$ for a cluster at redshift $z=0.2$, after cleaning the contamination from  CMB, synchrotron, dust using the Internal Linear Combination (ILC) method in the presence of the instrument noise of CMB-S4 \cite{Abazajian:2016yjj, 2019arXiv190704473A} given in table \ref{tab:s4}.}\label{Fig:ALP-map-1}
\end{figure}

We obtain the photon-ALP distortion signal from the contaminated sky,  we apply ILC (Internal Linear Combination) method \cite{Tegmark:1995pn, 2003ApJS..148....1B, Eriksen:2004jg} on this simulated map ($S_{\nu_i}(\hat p)$)  to extract the ALP signal which obeys the spectrum $\mathbf{f}_{\gamma a}$ given in Fig.~\ref{Fig:Inu-nu}. 
The noise minimized sky map for the ALP signal (denoted by $\hat x_{\gamma a}$) can be obtained as \cite{Tegmark:1995pn, 2003ApJS..148....1B, Eriksen:2004jg}
\begin{equation}\label{ilc-1a}
\hat x^{\{Q, U\}}_{\gamma a}(\hat p)=  \mathbf{{\hat W}}_{\gamma a}^T \mathbf{S}^{\{Q, U\}}(\hat p),
\end{equation}
where $\mathbf{\hat W}_{\gamma a}(\nu)=  \mathbf{C}_S^{-1} \mathbf{f}_{\gamma a} (\mathbf{f}^T_{\gamma a} \mathbf{C}_S^{-1}\mathbf{f}_{\gamma a})^{-1}$ are the ILC weight factors and $\mathbf{C}_S$ is the covariance matrix obtained from this map (which includes CMB, ALP signal, synchrotron, thermal dust and instrument noise). The maps of the ALP signal (for axion mass $10^{-13}$ eV and $g_{\gamma a}= 10^{-12}$ GeV$^{-1}$) around a single galaxy cluster at a redshift $z=0.2$ after performing the ILC cleaning is shown in Fig.~\ref{Fig:ALP-map-1}. ILC method successfully removes the foreground contamination and makes the photon-ALP distortion signal more evident in the sky, as can be understood by comparing Fig.~\ref{Fig:ALP-map-1-impure} and Fig.~\ref{Fig:ALP-map-1}.

\begin{figure}[H]
\centering
\begin{subfigure}{1\textwidth}
\centering
\includegraphics[trim={0 0 0 0cm}, clip, width=0.9\textwidth]{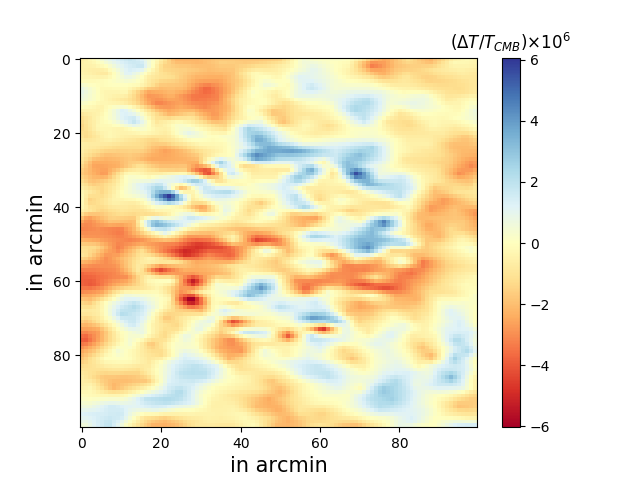}
 \caption{Q map of ALPs-distortion for single galaxy cluster located at redshift $z=0.1$.}\label{Fig:ALP-map-z1}
     \end{subfigure}
     \begin{subfigure}{1\textwidth}
     \centering
\includegraphics[trim={0 0 0 0cm}, clip, width=0.9\textwidth]{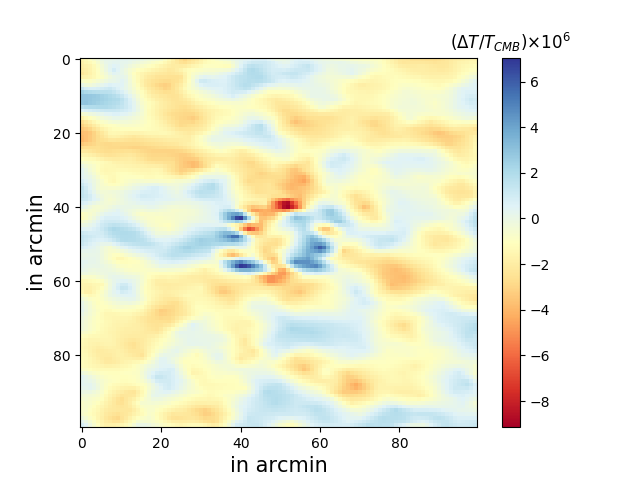}
 \caption{Q map of ALPs-distortion for single galaxy cluster located at redshift $z=0.3$. }\label{Fig:ALP-map-z3}
     \end{subfigure}
        \caption{ALPs-distortions around a single galaxy cluster for axion mass $m_a= 10^{-13}$ eV and photon-ALPs coupling $g_{\gamma a}= 10^{-12}$ GeV$^{-1}$ after cleaning the foreground contamination using the Internal Linear Combination (ILC) method in the presence of the instrument noise of CMB-S4 \cite{Abazajian:2016yjj, 2019arXiv190704473A}. The angular size of the distortion is larger for a cluster at lower redshift than the cluster at high redshift.}\label{Fig:ALP-map-z}
\end{figure}

The variation of the photon-ALP distortion signal with the change in the redshift of the galaxy cluster are shown in Fig.~\ref{Fig:ALP-map-z1} and Fig.~\ref{Fig:ALP-map-z3} respectively for redshift $z= 0.1$ and $z= 1.33$ after cleaning the foreground contamination using ILC method. 
There are two major effects on the photon-ALP distortion with the change in redshift. First, the angular size of the disc increases with redshift, which can be seen from Fig.~\ref{Fig:ALP-map-1} and Fig.~\ref{Fig:ALP-map-z}.  Second, the signal strength increases with redshift according to the Eq.~\eqref{Eq:gad-2}. The changes in the signal strength are not very evident from the plots due to the contamination from foregrounds.

\section{Forecast for the measurement of the photon-ALP distortion signal from Simons Observatory and CMB-S4}\label{ALP-forecast}
In this section, we discuss the method to detect photon-ALP distortions from galaxy clusters and obtain the detection limits achievable from the upcoming  CMB experiments such as SO \cite{Ade:2018sbj} and CMB-S4 \cite{Abazajian:2016yjj, 2019arXiv190704473A} .  In order to identify the photon-ALP distortion from the polarization map, we identify the location of galaxy clusters in the polarization map using the cluster locations identified from the temperature map of CMB using the tSZ signal.
Then one can identify the same location of galaxy cluster in the foreground cleaned $Q$ and the $U$ map of CMB polarization, which can be obtained after implementing the ILC method on the data for the known frequency spectrum of the photon-ALP distortion as described in Sec.~\ref{ilc}.
  
\begin{figure}[h]
 \centering
 \includegraphics[trim={0 0 0 0cm}, clip, width=1\textwidth]{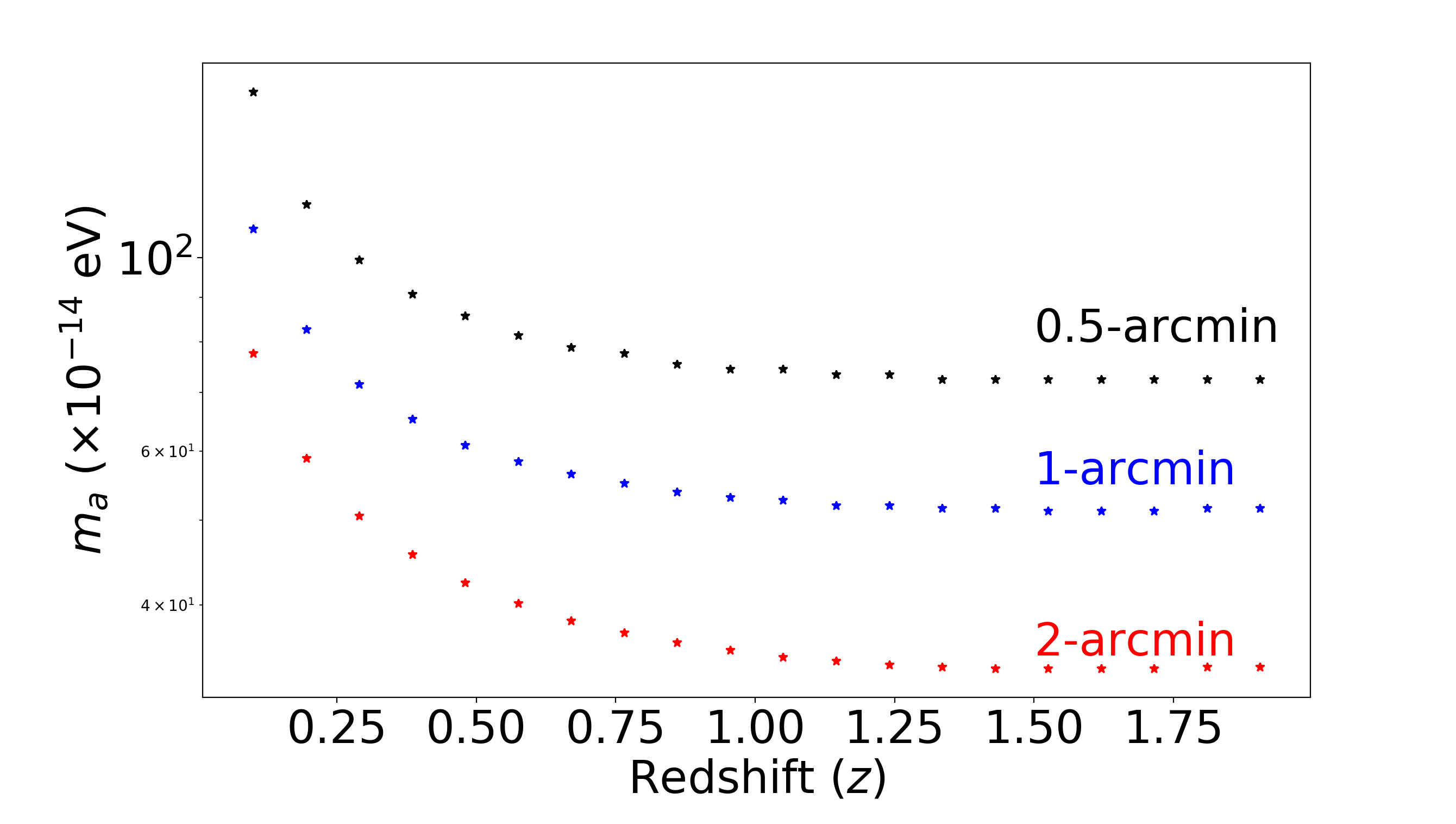}
          \captionsetup{singlelinecheck=on,justification=raggedright}
     \caption{The higher ALPs mass produces a smaller angular size at a fixed redshift. The smallest angular scale which can be resolved by an experiment depends upon the instrument beam. We plot here the maximum mass of ALPs as a function of redshift for which the distortion in the CMB polarization can be spatially resolved from the future CMB experiments having different beam resolutions such as 0.5 arcmin, 1 arcmin and 2 arcmin, and which obeys the electron density given in Eq. \ref{elec-den}.}\label{Fig:mass-z-res}
     \end{figure}

In order to measure the photon-ALP distortion for a particular ALP mass, we need to  resolve the disc of the distortion.  The spectral distortion signal from the large ALP mass originates at a smaller radius from the center of the galaxy cluster and hence can only be resolved from the low redshift objects, for a fixed instrument beam resolution. In Fig.~\ref{Fig:mass-z-res}, we show the maximum mass of the ALP as a function of the redshift which can be resolved by the CMB experiments having three different beam resolutions $0.5$ arcmin, $1$ arcmin and $2$ arcmin. The masses above the line cannot be accessed from the redshift range considered in this analysis. The plot implies that a future CMB experiment such as CMB-S4 \cite{Abazajian:2016yjj, 2019arXiv190704473A}  with $1$ arcmin beam resolution can probe ALP mass up to $10^{-12}$ eV.  The results shown in  Fig.~\ref{Fig:mass-z-res} are obtained for the model of the electron density described in Sec.~\ref{elec-den}  and are going to vary for different model of electron density. In the next section this uncertainty will inform our choice of aperture photometry to forecast the signal strength detectable with future CMB data. 

\subsection{Aperture photometry}\label{app-photo}
    There are several filtering techniques such as matched filtering \cite{Herranz:2002kg, Melin:2006qq}, constrained realization \cite{Abrial:2008mz, Inoue:2008qf, 2013A&A...549A.111E}, and aperture photometry \cite{2015MNRAS.451.1606F} that can be applied to recover the photon-ALP distortion signal around the clusters. Aperture photometry is more robust to the details of profile shape but less optimal than the matched filter approach, which relies on having a good, if not perfect, estimate of the  spatial shape of the photon-ALP distortion. With the improvement of our knowledge regarding the electron density and magnetic field of galaxy clusters, we can accurately estimate spatial profile of the photon-ALP distortion. This will make it possible to implement matched filtering technique to extract the photon-ALP distortion signal, which is going to improve SNR over the aperture photometry method. 
        
  \begin{figure}[h]
 \centering
 \includegraphics[trim={7cm 5cm 7cm 7cm}, clip, width=1\textwidth]{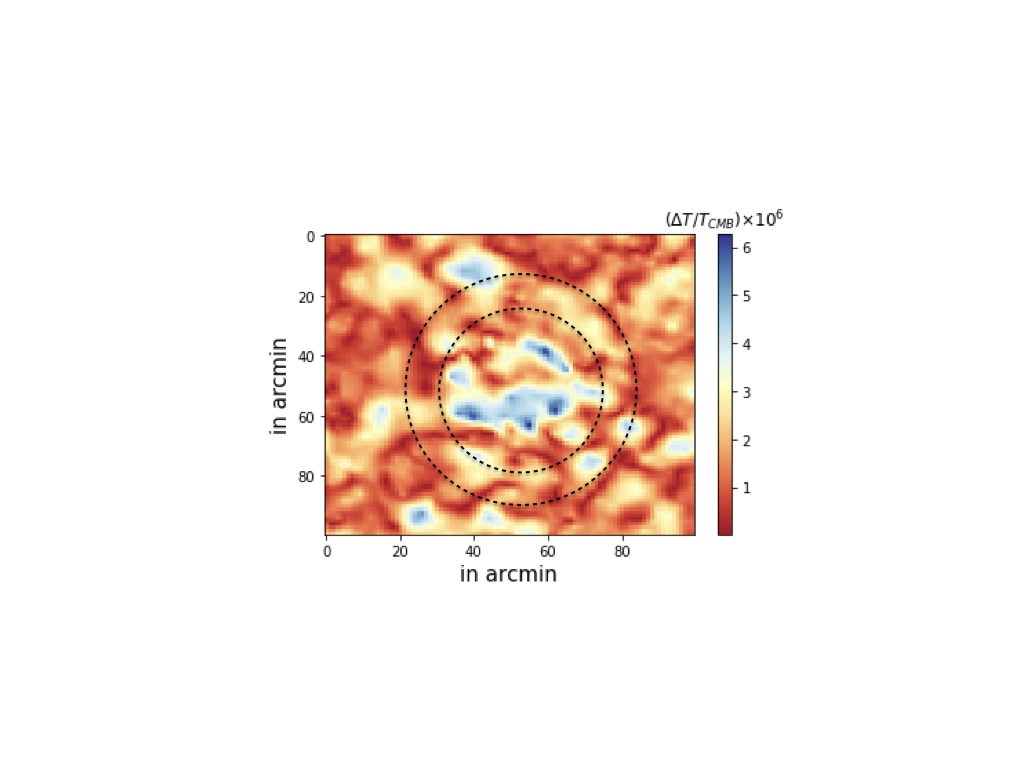}
          \captionsetup{singlelinecheck=on,justification=raggedright}
     \caption{The implementation of aperture photometry is shown in this diagram. The polarization intensity of the photon-ALP distortion $s^{\{P\}}_{\gamma a} = \sqrt{(\hat{x}^{\{Q\}}_{\gamma a})^2+ (\hat{x}^{\{U\}}_{\gamma a})^2}$ at a single cluster location is shown along with two discs of radius $\theta_d$ and  $\sqrt{2}\theta_d$. This plot is made for the ALP mass $m_a=10^{-13}$ eV and photon-ALPs coupling $g_{\gammaa}= 2\times 10^{-12}$ GeV$^{-1}$.}\label{Fig:ap-diagram}
     \end{figure}
Aperture photometry is a powerful filtering technique to extract the signal within a disc radius $\theta_d$ at a point in comparison to the neighboring sky area of equal area, and has been used on the temperature map \cite{2015MNRAS.451.1606F}. For the polarization map $Q$ and $U$, we can implement aperture photometry as
\begin{align}\label{ap-1}
\begin{split}
{AP_{\{P\}_{\gamma a}, \theta_d}= \int s^{\{P\}}_{\gamma a} (\theta) W_{\theta_d} (\theta) d^2\theta,}
\end{split}
\end{align}
where $s^{\{P\}}_{\gamma a} = \sqrt{(\hat{x}^{\{Q\}}_{\gamma a})^2+ (\hat{x}^{\{U\}}_{\gamma a})^2}$ is the polarization intensity of the foreground cleaned ALPs distortion at the cluster locations obtained from $\hat{x}^{\{Q,U\}}_{\gamma a}$ given in Eq.~\eqref{ilc-1a}. The window function $W_{\theta_d} (\theta)$ is defined as 
\begin{align}\label{window-1}
W_{\theta_d} (\theta)= \frac{1}{\pi \theta_d^2} 
\begin{dcases}
    1,& \text{if } \theta \leq \theta_d\\
    -1,& \text{if } \sqrt{2} \theta_d \geq \theta > \theta_d\\
    0             & \text{otherwise}.
\end{dcases}
\end{align}
This filtering scheme ensures the equal area of the disc and the neighboring circular annulus. The application of the window function projected on the polarization map is shown in Fig.~\ref{Fig:ap-diagram}. The total ALPs distortion intensity $I_{\gamma a, \theta_d}$ is the contribution from both the $Q$ and $U$ field which is given by the signal estimated using aperture photometry $AP_{\gamma a, \theta_d}$. The signal within the disc contains both the ALPs distortion and the background signal, whereas the outer disc of radius $\sqrt{2}\theta_d$ contains the polarization intensity only from the background fluctuations. Using the window function mentioned in Eq.~\eqref{window-1}, we subtract the background intensity of the polarization field from the inner disc of radius $\theta_d$. The covariance for the aperture photometry method can be written as

\begin{align}\label{ap-2}
\begin{split}
\mathcal{C}_{\theta_d, \theta'_{d}} \equiv \langle AP_{\gamma a, \theta'_d} AP_{\gamma a, \theta_{d'}} \rangle =  \int \frac{d^2l}{(2\pi)^2} W_{\theta_d}(l)W_{\theta'_{d}}(l') \mathcal{N}_l^{\{P\}_{ILC}},
\end{split}
\end{align}
where $\mathcal{N}_l^{\{P\}_{ILC}}$ is the power spectrum of the ILC polarization intensity map and the $W_{\theta_d}(l)$ is the Fourier transform of the window function $W_{\theta_{d}(\theta)}$. For a fixed ALP mass at a given fixed redshift, the signals will originate from a particular $\bar \theta_d$ at a fixed redshift of the galaxy cluster. So in this analysis, we will estimate the signal for the diagonal covariance matrix. 

For the implementation of this method on data of CMB-S4, we also need to mask the center part of the galaxy cluster, in order to remove the contamination near the core of the galaxy cluster from the secondary polarization anisotopies due to the scattering of the local CMB quadrupole with the electrons in the galaxy cluster, $\mathcal{P}_Q$ (See Eq. \eqref{optical_depth_cl}). In that case, we need to implement aperture photometry over a ring instead of a disc and estimate the covariance matrix for the corresponding window function. In this analysis, we have not included $\mathcal{P}_Q$ in simulated the sky signal. So our forecast presented here are not biased by the contamination from $\mathcal{P}_Q$. 

\subsection{SNR to measure the photon-ALPs distortion from SO and CMB-S4}
\begin{figure}[h]
 \centering
     \includegraphics[trim={0 0 0 0cm}, clip, width=1.\textwidth]{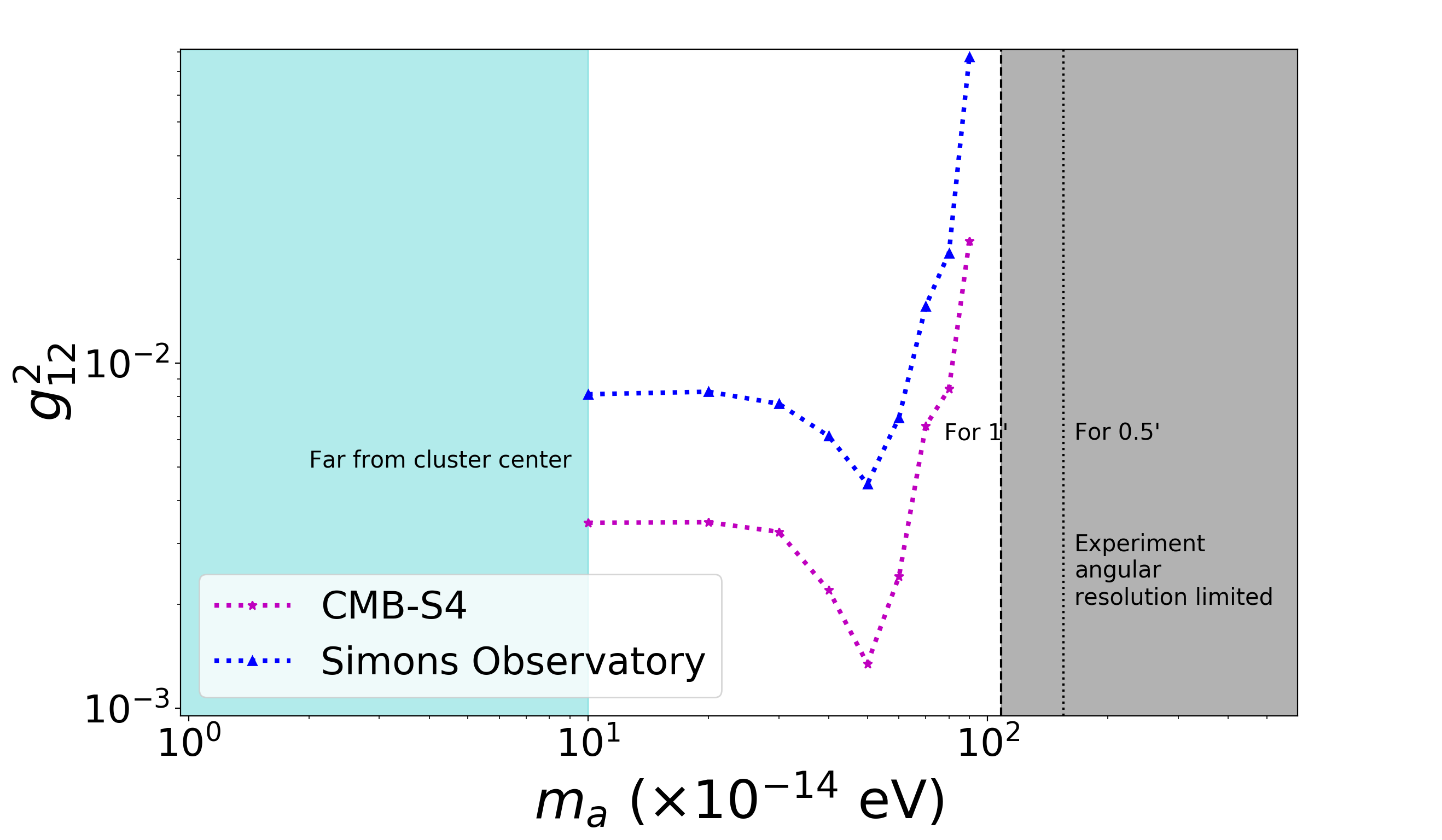}
          \captionsetup{singlelinecheck=on,justification=raggedright}
     \caption{$1\sigma$ error bar achievable from CMB-S4 and Simons Observatory on the photon-ALP coupling strength $g_{12} \equiv (g_{\gamma a} $ in units of $10^{-12}$ GeV$^{-1})$ are shown as a function of ALP masses. The fiducial value of the magnetic field is taken according to the model given in Eq.~\eqref{mag-field} and the fiducial value of the photon-ALP coupling $g_\gammaa$ is taken as zero. The black dashed line and dotted line shows the mass of the ALPs for which we can resolve the spectral distortions in the CMB with an instrument beam of 1 arc-minute and 0.5 arc-minute respectively.}\label{Fig:mass-g-snr}
     \end{figure}

Future large aperture CMB polarization experiments such as SO \cite{Ade:2018sbj} and CMB-S4 \cite{Abazajian:2016yjj, 2019arXiv190704473A}, are capable to measure clusters up to a redshift $z=2$. For this analysis, we have taken the cluster number per redshift bin ($N(z)/(\Delta z=0.1)$) as shown in Fig.~\ref{Fig:dn-dz}, which forecast for  SO   \cite{Ade:2018sbj} and CMB-S4 \cite{Abazajian:2016yjj, 2019arXiv190704473A}. Using this $N(z)/\Delta z$, we estimate the total SNR by filtering the distortion signal within a disc size $\theta_d$ at every cluster location by using the aperture photometry method. The total SNR added over all the redshifts and objects can be written as
\begin{align}\label{snr-1}
\begin{split}
\bigg(\frac{S}{N}\bigg)_{m_a}^2 = \sum_z\sum_{N(z)} \frac{(AP_{\gamma a, \theta_d(m_a,z)})^2}{\mathcal{C}_{\theta_d(m_a,z), \theta_{d}(m_a,z)}}.
\end{split}
\end{align}
The above equation for the SNR includes all other contamination such as CMB, synchrotron, dust and the instrument noise (SO \cite{Ade:2018sbj} or  CMB-S4 \cite{Abazajian:2016yjj, 2019arXiv190704473A}) through the covariance term $\mathcal{C}^I_{\theta_d, \theta'_{d}}$. Using the $N(z)$ (plotted in Fig.~\ref{Fig:dn-dz}), we show the $1$-$\sigma$ error bar on the photon-ALP coupling strength $g_\gammaa$ for different masses of ALPs in Fig.~\ref{Fig:mass-g-snr}, after marginalizing over the strength of magnetic field with a $30\%$ prior on its amplitude given in Eq. \ref{mag-field}. The region shaded in cyan color indicates the part of the parameter space which can be explored with a better knowledge of the electron density and magnetic field at distances farther away from the core of the galaxy cluster.  In this analysis, we consider the maximum radius of around $4$ Mpc from the cluster center, which leads to a minimum ALP mass $10^{-13}$ eV. The region shaded in grey indicates the inaccessible parameter space due to the limitation from the angular resolution of the instrument beam. The maximum mass of ALPs which can be probed by a $1$ arcmin and $0.5$ arcmin instrument beam are indicated by the black dashed line and dotted line respectively. The SNR is going to improve with a better knowledge of the electron density of galaxy clusters. This is going to be  possible from the tSZ measurement from the CMB temperature map of SO \cite{Ade:2018sbj} and CMB-S4 \cite{Abazajian:2016yjj, 2019arXiv190704473A}, as well as from the X-ray observations from the mission such as eROSITA \cite{2012arXiv1209.3114M}. Measurement of the cluster magnetic field from SKA \cite{Loi_2017} is also going to improve the projected SNR. By including the galaxy clusters catalogue from the upcoming large scale structure surveys \citep{Aghamousa:2016zmz, 2010arXiv1001.0061R, 2009arXiv0912.0201L,Dore:2018kgp,Dore:2018smn}, we are going to make further improvement in the measurement of the photon-ALPs coupling strength. 

The discovery space of this approach to detect ALPs is going to be broadened with a very high resolution future CMB mission such as CMB-HD \cite{Sehgal:2019ewc}. CMB-HD proposes to measure the microwave sky over the frequency bands $30-280$ GHz, with an antenna of diameter $30$ meters, resulting into an angular resolution of $\sim 15$ arcseconds at $\nu= 150$ GHz and better sensitivity ($0.7$\, $\mu$K- arcminute at $\nu= 150$ GHz for polarization). Such a mission is going to improve the feasibility of detecting ALPs in two ways: (i) by exploring weaker photon-ALP coupling strength $g_{\gammaa}$, and (ii) larger masses of ALPs due to better angular resolution of the instrument beam.  This will enable us to detect photon-ALP coupling up to higher ALPs mass $m_a\sim 2\times 10^{-12}$ eV. Assuming a large number of galaxy clusters ($10 \times$ CMB-S4) up to the cluster masses $\sim 10^{12}\, M_\odot$ from CMB-HD, the measurement of the photon-ALP coupling strength $g_{\gammaa}$ would result in  an error bar about three times smaller than CMB-S4 \cite{Abazajian:2016yjj, 2019arXiv190704473A} for  ALPs masses in the range $m_a \sim 10^{-13}$- $10^{-12}$ eV.

 \begin{figure}[h]
 \centering
     \includegraphics[trim={0 0 0 0cm}, clip, width=1.\textwidth]{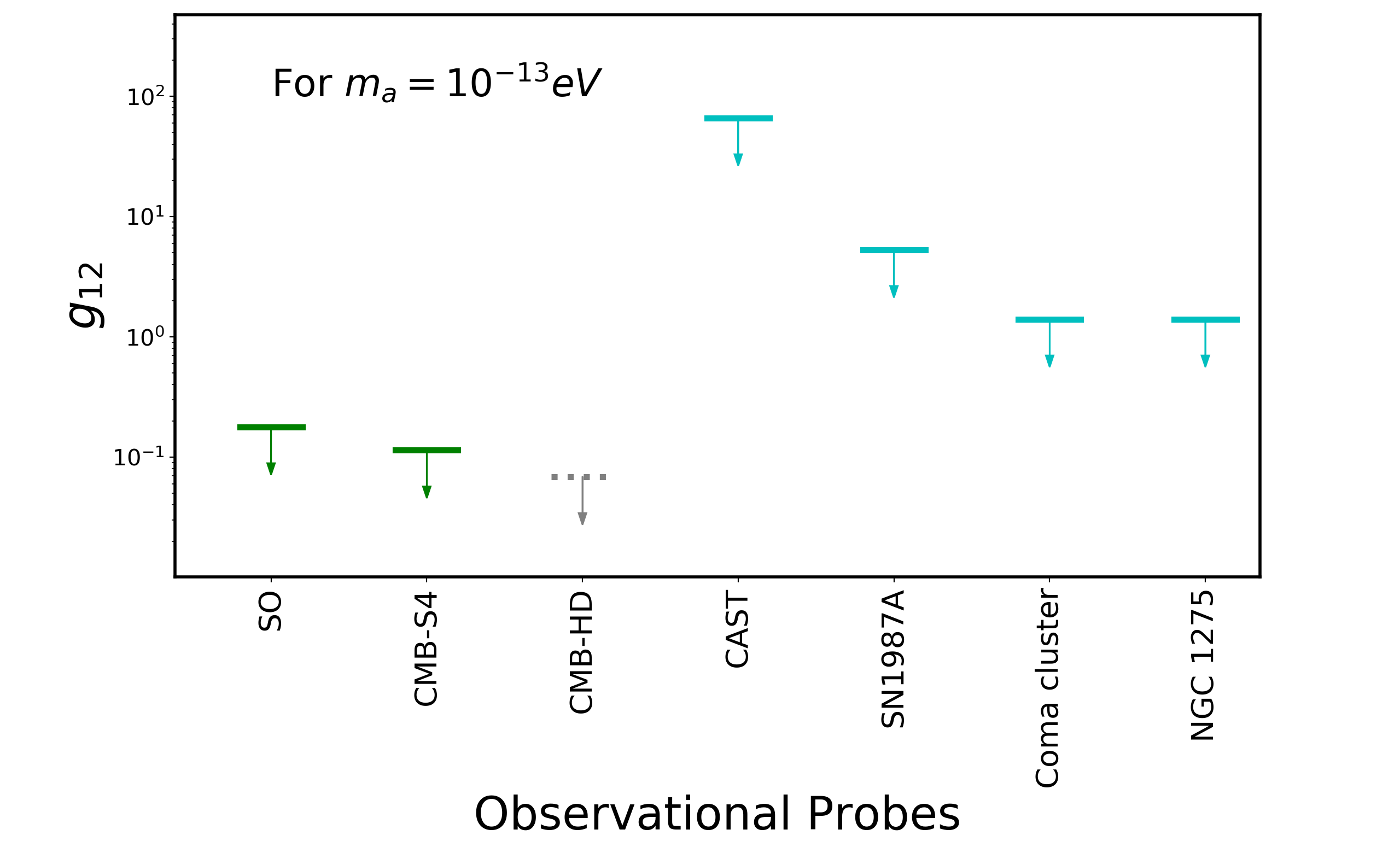}
          \captionsetup{singlelinecheck=on,justification=raggedright}
     \caption{The projected $2$-$\sigma$ constraint on the photon-ALPs coupling strength $g_{\gammaa}$ ($g_{12} \equiv (g_{\gamma a} $ in units of $10^{-12}$ GeV$^{-1})$) from the upcoming CMB experiments such as SO \cite{Ade:2018sbj}, and CMB-S4 \cite{Abazajian:2016yjj, 2019arXiv190704473A} are compared with the currently existing bounds from CAST \cite{Anastassopoulos:2017ftl}, SN1987A \cite{2015JCAP...02..006P}, X-ray observations from the Coma cluster \cite{Conlon:2015uwa} and NGC1275 \cite{Berg:2016ese}. Scaling the CMB-S4 constraints for $10$ times more galaxy cluster gives the constraint labelled CMB-HD \cite{Sehgal:2019ewc} (grey).}\label{Fig:mass-g-missions}
     \end{figure}
     
The bound on photon-ALP coupling $g_\gammaa$ possible from SO  \cite{Ade:2018sbj} and CMB-S4 \cite{Abazajian:2016yjj, 2019arXiv190704473A} are about two orders of magnitude better than the current bound $g_{\gammaa} \leq 6.6 \times 10^{-11}$ GeV$^{-1}$ and $g_{\gammaa} \leq 5.3 \times 10^{-12}$ GeV$^{-1}$ available from the CERN ALP Solar Telescope (CAST) \cite{Anastassopoulos:2017ftl} and  SN1987A \cite{2015JCAP...02..006P} respectively.  {Stringent} bound on the photon-ALPs coupling as $g_\gammaa \leq 1.4 \times 10^{-12}$ GeV$^{-1}$ was obtained using the X-ray observation of Coma cluster \cite{Conlon:2015uwa} and NGC1275 \cite{Berg:2016ese}. However, this bound can get weaker due to uncertainty in the strength of the cluster magnetic field. A plot comparing the possible constraints on the photon-ALPs coupling strength $g_\gammaa$ from the upcoming CMB experiments with the existing bounds from CAST \cite{Anastassopoulos:2017ftl} and SN-1987A \cite{2015JCAP...02..006P} are shown in Fig. \ref{Fig:mass-g-missions}.  {These results are obtained assuming that the strength of the coherent magnetic field have a radial dependence mentioned in Sec. \ref{ALP-cluster} and the direction of the magnetic field is randomly oriented in galaxy cluster. However the impact of turbulence within the spatial shell where resonance conversion takes place is ignored in this analysis. A detailed analysis including the variation of the photon-ALPs signal for different electron density and galaxy magnetic field including the turbulent component will be addressed in a future analysis.}

\section{Conclusions and future prospects}\label{ALP-conclusion}
Analogous to the tSZ spectral distortion due to inverse-Compton scattering in the CMB temperature signal around the galaxy clusters, ALPs can also produce spectral distortions in the polarization field of CMB photons around the galaxy clusters due to the resonant conversion from photons to ALPs in the presence of the magnetic field of galaxy cluster. The ALP of different masses can undergo resonant conversion at different radii from the center of the cluster where the mass of ALP equals the photon mass in the plasma. As a result, the photon-ALP distortion around the galaxy cluster depends upon the spatial structure of the electron density in the cluster. For a universal spherical electron density in the galaxy cluster, the signal appears in a disk-shaped region for every galaxy cluster. Deviation from the spherical electron density profile will exhibit a non-circular shape around the galaxy cluster. In a future analysis, we will study the variation of this ALPs distortion signal for different electron density and magnetic field models and will explore the robustness of this feature. We  estimate the photon-ALP spectral distortion signal which can originate from galaxy cluster for the model of electron density and the magnetic field as mentioned in Eq. \ref{elec-den} and Eq. \ref{mag-field} respectively. The photon-ALP distortion signal exhibits higher strength for lower ALPs mass and the signal appears in the sky at a larger radius as shown in Fig.~\ref{Fig:pgamma_x}. The conversion probability depends on the frequency of the CMB photons at the frame of the galaxy cluster and as a result depends on the redshift of the galaxy cluster as shown in Fig.~\ref{Fig:mass-r-1}. The frequency dependence of photon-ALP distortion signal is different from the other spectral distortion signals and known foreground contamination, as shown in Fig .\ref{Fig:Inu-nu}. This makes it possible  to distinguish the photon-ALP distortion signal from other signals in the sky. 

In order to simulate a realistic sky signal, we simulate polarized foregrounds for the synchrotron, dust using the numerical code PySM \cite{Thorne:2016ifb},  along with CMB fluctuations and the instrument noise for SO \cite{Ade:2018sbj} and CMB-S4 \cite{Abazajian:2016yjj, 2019arXiv190704473A}. The spectral distortion signal from photon-ALP around the galaxy clusters are added at randomly chosen cluster locations, outside the plane of the galaxy clusters. The spatial profile of the photon-ALP distortion signal around a galaxy clusters are shown in Sec.~\ref{ALP-contamination}. We implement the internal linear combination (ILC) method to extract the photon-ALP distortion signal from the simulated data, and have shown the ILC cleaned photon-ALP distortion signal for different galaxy clusters at redshift $z=0.1,\,0.2, \,1.33$ in Fig.~\ref{Fig:ALP-map-1}, Fig .\ref{Fig:ALP-map-z1} and Fig.~\ref{Fig:ALP-map-z3} respectively. 

Using the aperture photometry technique, as discussed in Sec.~\ref{app-photo}, we filter out the non-ALP background distortion signal and estimate the polarization intensity of the photon-ALP distortion signal using the ILC cleaned polarization maps. For the redshift distribution of the cluster catalog (as shown in Fig.~\ref{Fig:dn-dz}) which are expected from the CMB experiments such as SO \cite{Ade:2018sbj} and CMB-S4 \cite{Abazajian:2016yjj, 2019arXiv190704473A}, we show the $1$-$\sigma$ error bar on the photon-ALP coupling strength $g_\gammaa$ for SO \cite{Ade:2018sbj} and CMB-S4 \cite{Abazajian:2016yjj, 2019arXiv190704473A} in Fig.~\ref{Fig:mass-g-snr}.  {These results are obtained for a particular radial dependence of the cluster magnetic field with random directions of the magnetic field. In a future work, we will study the impact of different magnetic field and electron density models in the present of turbulence.} This new avenue is going to explore new parameter space of photon-ALP coupling which is beyond the reach of the ground-based particle physics experiment such as CERN ALP Solar Telescope (CAST) \cite{Anastassopoulos:2017ftl} for ALP  masses in the range $10^{-13}-10^{-12}$ eV. ALPs of lower masses (up to $\sim 10^{-13}$ eV) can be probed by using the measurement of the magnetic field of galaxy clusters from the upcoming experiment such as Square Kilometer Array (SKA) \cite{Loi_2017}. The synergy between SO \cite{Ade:2018sbj}, CMB-S4 \cite{Abazajian:2016yjj, 2019arXiv190704473A} and SKA \cite{Loi_2017} can be used to explore a broader parameter space of ALPs. With a future CMB experiment such as CMB-HD \cite{Sehgal:2019ewc}, we can also extend the reach of our method to higher ALPs masses than what can be possible from SO \cite{Ade:2018sbj} and CMB-S4 \cite{Abazajian:2016yjj, 2019arXiv190704473A}. We can probe the photon-ALP coupling strength $g_\gammaa$ for $m_a> 10^{-12}$ eV from CMB-HD \cite{Sehgal:2019ewc}.  

 {One of the contaminations} to the ALPs distortion will come from the scattering of the local quadrupole by the hot electrons in the galaxy cluster ($\mathcal{P}_Q$) which can shadow the photon-ALP spectral distortion signal for the coupling strength $g_\gamma=10^{-14}$ GeV$^{-1}$ for $r<3500$kpc, as discussed in Sec.~\ref{ALP:singlecluster}. For the $1$-$\sigma$ error-bars on $g_\gammaa$ from SO \cite{Ade:2018sbj} and CMB-S4 \cite{Abazajian:2016yjj, 2019arXiv190704473A}, as shown in Fig.~\ref{Fig:mass-g-snr}), the bias from $\mathcal{P}_Q$ can affect the CMB-S4 measurements by about a $1$-$\sigma$ for CMB-S4 \cite{Abazajian:2016yjj, 2019arXiv190704473A}  and only by about one-third of a $\sigma$ for SO  \cite{Ade:2018sbj}. A joint estimator of both photon-ALP distortion and $\mathcal{P}_Q$ needs to be developed for CMB-S4 \cite{Abazajian:2016yjj, 2019arXiv190704473A}  like survey in order to make an unbiased estimation of both $\mathcal{P}_Q$ and $P_{\gammaa}$. This will be addressed in a future analysis. 

This new powerful avenue to probe ALPs can also be applied to the currently detected galaxy clusters from ACTPol \cite{Henderson:2015nzj}, SPT \cite{Benson:2014qhw}, and Planck \cite{2016A&A...594A...1P}. We leave this work for a future publication. The ALPs signal from the galaxy clusters can also be probed using photons in other wavebands, such as the X-ray galaxy catalogs from Chandra X-ray Observatory \cite{2014RScI...85f1101S} and  from the recently launched eROSITA misssion \cite{2012arXiv1209.3114M}. In summary, CMB polarization is a powerful probe of ALPs physics and  should be explored in parallel with ongoing laboratory experiments such as CAST.

\textbf{Acknowledgement}
SM would like to thank Daniel Baumann, Daniel Green, Renee Hlozek, Lyman Page and Joseph Silk for useful discussions. SM would also like to thank Colin Hill for providing the details of the instrument noise for Simons Observatory. SM acknowledge the use of PySM package \cite{Thorne:2016ifb} and would like to thank Jo Dunkley for mentioning about this package. The Flatiron Institute is supported by the Simons Foundation. SM, DNS and BDW acknowledge the support of the Simons Foundation for this work. The work of SM and BDW is also supported by the Labex ILP (reference ANR-10-LABX-63) part of the Idex SUPER, and received financial state aid managed by the Agence Nationale de la Recherche, as part of the programme Investissements d'avenir under the reference ANR-11-IDEX-0004-02. BDW is supported through the ANR BIG4 project under reference ANR-16-CE23-0002. RK was supported by SERB grant no. ECR/2015/000078 from Science and Engineering Research Board, Department of Science and Technology, Govt. of India and MPG-DST partner group between Max Planck Institute for Astrophysics, Garching and Tata Institute for Fundamental Research, Mumbai funded by Max Planck Gesellschaft. The computational works are performed in the Rusty cluster of the Flatiron Institute. A part of the computational work is also done in the Horizon Cluster hosted by Institut d'Astrophysique de Paris. We thank Stephane Rouberol for running smoothly this cluster for us. We acknowledge the use of following packages in this analysis: Astropy \cite{2013A&A...558A..33A, 2018AJ....156..123A}, IPython \cite{PER-GRA:2007}, HEALPix (Hierarchical Equal Area isoLatitude Pixelation of a sphere)  \footnote{Link to the HEALPix website \url{https://healpix.sourceforge.io}} \cite{Gorski:2004by}, Mathematica \cite{mathematica}, Matplotlib \cite{Hunter:2007},  NumPy \cite{2011CSE....13b..22V}, and SciPy \cite{scipy}.

\def\urlprefix{}
\def\url#1{}
\bibliography{cluster-axion}
\bibliographystyle{JHEP}
\end{document}